\documentclass{aa}  

\usepackage{graphicx}
\usepackage{natbib}
\usepackage{array,multirow,makecell} 
\usepackage{amssymb}
\usepackage{amsmath}
\usepackage{pifont}
\usepackage{indentfirst}
\usepackage[breaklinks, colorlinks, citecolor=blue]{hyperref}
\usepackage[dvipsnames]{xcolor}

\usepackage[varg]{txfonts}
\usepackage{bm}

\def\drm{\mathrm{d}}
\def\los{{\bf \hat n}}
\def\vvec{{\bf v}}
\def\zrei{{z_\mathrm{re}}}
\def\zend{{z_\mathrm{end}}}
\def\fesc{{f_\mathrm{esc}}}
\def\lmax{{\ell^\mathrm{max}}}
\def\xhii{{x_e}}

\def\kvec{{\bf  k}}
\def\kvecunit{{\bf \hat k}}
\def\Pdd{P_{\delta \delta}}
\def\Plin{P_{\delta \delta}^\mathrm{lin}}
\def\Pee{P_{ee}}
\def\dthree{\mathcal{D}_{3000}}
\def\qvec{{\bf \tilde {q}}}
\def\qvecreal{{\bf {q}}}
\def\exp{{\rm e}}

\begin{document} 

\title{Improved constraints on reionisation from CMB observations:\\A parameterisation of the kSZ effect}

\author{
    A. Gorce \inst{1,2}
    \and S. Ili\'c \inst{3,4,5}
    \and M. Douspis \inst{1}
    \and D. Aubert \inst{6}
    \and M. Langer \inst{1}
}

\institute{
    Université Paris-Saclay, CNRS,  
    Institut d'Astrophysique Spatiale, 
    91405, Orsay, France\\
	\email{adelie.gorce@ias.u-psud.fr}
	\and
	Department of Physics,
	Blackett Laboratory, 
	Imperial College,			
	London SW7 2AZ, U.K.
	\and
	CEICO, Institute of Physics of the Czech Academy of Sciences,
	Na Slovance 2, Praha 8, Czech Republic
	\and
	Universit\'e PSL, Observatoire de Paris, Sorbonne Universit\'e, CNRS, LERMA, F-75014, Paris, France
	\and
	IRAP, Universit\'e de Toulouse, CNRS, CNES, UPS, Toulouse, France
	\and
	Observatoire Astronomique de Strasbourg,
	Université de Strasbourg,
	CNRS UMR 7550,
	11 rue de l’Université,
	67000 Strasbourg, France
}

\date{Received *********; accepted *******}

\abstract{We show that, in the context of patchy reionisation, an accurate description of the angular power spectrum of the kinetic Sunyaev-Zel'dovich (kSZ) effect is not possible with simple scaling relations between the amplitude of the spectrum and global parameters, such as the reionisation midpoint and its duration. We introduce a new parameterisation of this spectrum, based on a novel description of the power spectrum of the free electrons density contrast $\Pee (k,z)$ in terms of the reionisation global history and morphology. We directly relate features of the spectrum to the typical ionised bubble size at different stages in the process and, subsequently, to the angular scale at which the patchy kSZ power spectrum reaches its maximum. We successfully calibrated our results on a custom set of advanced radiative hydrodynamical simulations and later found our parameterisation to be a valid description of a wide range of other simulations and, therefore, reionisation physics. In the end, and as long as the global reionisation history is known, two parameters are sufficient to derive the angular power spectrum. Such an innovative framework applied to cosmic microwave background data and combined with 21cm intensity mapping will allow a first consistent detection of the amplitude and shape of the patchy kSZ signal, giving in turn access to the physics of early light sources.}
\keywords{Cosmology: dark ages, reionization, first stars -- cosmic background radiation -- Methods: analytical}

\maketitle

\section{Introduction}

From the launch of the Cosmic Background Explorer (COBE) in 1989 to the publication of the latest results of the Planck satellite in 2018 \citep{planck_2018_overview}, the study of the cosmic microwave background (CMB) has triggered a tremendous amount of research. Cosmological parameters have been estimated with exquisite precision and our knowledge of cosmic inflation has been greatly improved. Along the line of sight, the primordial part of the CMB signal is largely modified by the interaction of CMB photons with structures that formed later in the Universe. Notably, their interaction with free electrons in the intergalactic medium (IGM)  modify the shape and amplitude of the measured CMB temperature and polarisation power spectra. The presence of these electrons is the result,  in particular, of cosmic reionisation, an era potentially extending from a redshift of $z \sim 15$ to $z \sim 5$ when the first galaxies are thought to have ionised the neutral hydrogen and helium in the surrounding IGM. 

CMB photons  lose energy from scattering off low-energy electrons. In CMB data analysis, this effect is accounted for when computing the Thomson optical depth. To do so, one needs to assume a global history of reionisation, that is, a redshift-evolution for the IGM global ionised fraction $x_e (z)$. In standard Boltzmann solvers which are used to compute theoretical predictions in CMB data analysis such as the \texttt{CAMB} code (\mbox{\citealt{camb1}}, \citealt{camb2})\footnote{Available at \url{https://camb.info}.}, the reionisation scenario used is a step-like transition of $x_e(z)$, where the global ionised fraction jumps from $10 \%$ to $75 \%$ over a (fixed) redshift interval of $\Delta z = 1.73$ \citep{planck_2015_overview}. However, this parameterisation does not match simulations and observations well since we expect the ionisation fraction to slowly rise when the first sources light up, before taking off as soon as about $20\%$ of the IGM is ionised (\mbox{\citealt{Robertson_2015}}, \citealt{greig_mesinger_2016}, \citealt{gorce_2018}). This minimal model can have a huge impact on reionisation constraints: The value of $\tau$ inferred from Planck 2016 data varies from $0.066 \pm 0.016$ for a step-like process to $0.058 \pm 0.012$ for a more accurate description \citep{douspis_2015,planck_2016_reio}. It is therefore essential to take the asymmetric evolution of $x_e(z)$ into account when trying to accurately constrain reionisation, and global parameters such as the reionisation midpoint $\zrei$ and duration $\Delta z$ are not sufficient.

CMB photons can also gain energy from scattering off electrons with a non-zero bulk velocity relative to the CMB rest-frame in a process called the kinetic Sunyaev-Zel'dovich effect \citep[hereafter kSZ effect, see][]{zeldovich_sunyaev_1969,sunyaev_zeldovich_1980}. This interaction adds power to the CMB temperature spectrum on small angular scales ($\ell \gtrsim 2000$, that is smaller than about 5 arcminutes), where secondary anisotropies, including kSZ, dominate the signal. The impact of kSZ on the CMB power spectrum is often split between the homogeneous kSZ signal, which come from the Doppler shifting of photons on free electrons that are homogeneously distributed throughout the IGM once reionisation is over, and the patchy kSZ signal, when CMB photons scatter off isolated ionised bubbles along the otherwise neutral line of sight. Therefore, the kSZ power spectrum is sensitive to the duration and morphology of reionisation (\mbox{\citealt{mcquinn_2005}}, \citealt{mesinger_2012_kSZ}). For example, the patchy signal is expected to peak around $\ell \sim 2000$, corresponding to the typical bubble size during reionisation \citep{zahn_2005,iliev_2007}. 

Secondary anisotropies only dominate the primordial power spectrum on small scales, where existing all-sky surveys such as Planck perform poorly. The observational efforts of the ground-based Atacama cosmology telescope (ACT)\footnote{\url{https://act.princeton.edu}} and the South Pole telescope (SPT)\footnote{\url{http://pole.uchicago.edu}} have allowed upper constraints to be put on the amplitude of the kSZ power spectrum at $\ell = 3000$. Using ACT observations at $148~\mathrm{GHz}$, \citet{dunkley_2011_act} find $\dthree^\mathrm{SZ} \equiv \ell \left( \ell + 1 \right) C_{\ell=3000}^\mathrm{SZ}/2\pi = 6.8 \pm 2.9 ~ \mu \mathrm{K}^2$ at the $68\%$ confidence level (C.L.) for the sum of thermal and kinetic SZ. In a first analysis, \citet{reichardt_2012_spt} derive from the three frequency bands used by SPT $\dthree^\mathrm{kSZ} < 2.8 ~ \mu \mathrm{K}^2$ ($95\%$ C.L.). This limit is however significantly loosened when anti-correlations between the thermal SZ effect (tSZ) and the cosmic infrared background (CIB) are considered. By combining SPT results with large-scale CMB polarisation measurements, \citet{zahn_2012_spt} are subsequently able to constrain the amplitude of the patchy kSZ by setting an upper limit $\dthree^\mathrm{patchy} \leq 2.1 ~ \mu\mathrm{K}^2$ ($95\%$ C.L.) translated into an upper limit on the duration of reionisation $\Delta z \equiv z\left(x_e=0.20\right) - z\left(x_e=0.99\right) \leq 4.4$ ($95\%$ C.L.), again largely loosened when CIB$\times$tSZ correlations are considered. Using Planck's large-scale temperature and polarisation ($EE$) data, combined with ACT and SPT high-$\ell$ measurements, and taking the aforementioned correlations into account, \citet{planck_2016_reio} find a more constraining upper limit on the total kSZ signal $\dthree^\mathrm{kSZ} < 2.6~\mu\mathrm{K}^2$ with a $95\%$ confidence level. Finally, adding new data from SPTpol\footnote{The second camera deployed on SPT, polarisation sensitive.}  to their previous results \citep{george_2015_spt}, \citet{SPT_2020} claim the first $3\sigma$ detection of the kSZ power spectrum with an amplitude $\dthree^{kSZ} = 3.0 \pm 1.0~\mu\mathrm{K}^2$, translated into a confidence interval on the patchy amplitude $\dthree^{\mathrm{p}kSZ}=1.1^{+1.0}_{-0.7}~\mu\mathrm{K}^2$ using the models of homogeneous signal given in \citet{shaw_2012}. These results are further pushed using the scaling relations derived by \citet{battaglia_2013_paperIII} to obtain an upper limit on the duration of reionisation $\Delta z < 4.1$.

Previous works have focused on relating the amplitude of the kSZ power spectrum at $\ell = 3000$ to common reionisation parameters such as its duration and its midpoint. \citet{battaglia_2013_paperIII} use large dark matter simulations ($L \gtrsim 2~\mathrm{Gpc}/h$), post-processed to include reionisation, to construct light-cones of the kSZ signal and estimate its patchy power spectrum. The authors find the scalings $\dthree^\mathrm{kSZ} \propto \bar{z}$ and $\dthree^\mathrm{kSZ} \propto \Delta z^{0.51}$ where $\bar{z}$ is approximately the midpoint of reionisation and here $\Delta z \equiv z\left(x_e=0.25\right) - z\left(x_e=0.75\right)$. Very large box sizes are necessary to capture the large-scale velocity flows contributing to the kSZ power spectrum at high-$\ell$ and results based on insufficiently large simulations will significantly underestimate the power at these scales. \citet{shaw_2012} find that a simulation box of side length $100~\mathrm{Mpc}/h$ would miss about $60\%$ of the kSZ power at $\ell = 3000$. For their own work, \citet{shaw_2012} therefore choose a completely different approach: they use hydrodynamical simulations to map the gas density to the dark matter power spectrum and later include this bias in a purely analytical derivation of the kSZ angular power spectrum. Because the non-linear dark matter power spectrum can be computed using the \texttt{HALOFIT} procedure \citep{smith_2003_halofit} and because the velocity modes can be estimated fully from linear theory under a few assumptions, they avoid the limitations caused by simulation resolution and size mentioned above. With this method, the authors find a power-law dependence on both the reionisation midpoint $\zrei$ and the optical depth $\tau$ for the homogeneous signal. For their most elaborate simulation, dubbed CSF, the cosmology-dependent scaling relations write $\dthree^{kSZ} \propto \tau^{0.44}$ and $\dthree^{kSZ} \propto \zrei^{0.64}$ but are independent since one parameter is fixed before varying the other. The authors note that the current uncertainties on cosmological parameters such as $\sigma_8$ will wash out any potential constraint on $\zrei$ and $\tau$ obtained from the measurement of the kSZ spectrum. 

In this work, we choose to follow a similar approach. We build a comprehensive parameterisation allowing the full derivation of the kSZ angular power spectrum from a known reionisation history and morphology. In Sec.~\ref{sec:equations}, we review the theoretical derivation of the kSZ power spectrum and propose a new parameterisation of the power spectrum of free electrons density contrast, based on the shape of the power spectrum of a bubble field. In Sec.~\ref{sec:calibration}, we present the simulations we later use to calibrate this parameterisation. In Sec.~\ref{sec:results}, we use the resulting expression of $\Pee(k,z)$ to compute the patchy kSZ angular power spectrum of our simulations and later apply the same procedure to different types of reionisation simulations. Finally, in Sec.~\ref{sec:conclusion}, we discuss the physical meaning of our parameters and conclude. All distances are in comoving units and the cosmology used is the best-fit cosmology derived from Planck 2015 CMB data \citep{planck_2015_overview}: $h = 0.6774$, $\Omega_\mathrm{m} = 0.309$, $\Omega_\mathrm{b} h^{2} = 0.02230$, $Y_\mathrm{p} = 0.2453$, $\sigma_8 =0.8164$ and $T_\mathrm{CMB} = 2.7255~\mathrm{K}$. Unless stated otherwise, $\Pdd$ describes the non-linear total matter power spectrum, $x_e(z)$ is the ratio of $\ion{H}{II}$ and $\ion{He}{II}$ ions to protons in the IGM, and the reionisation duration is defined by $\Delta z = z\left(x_e=0.25\right) - z\left(x_e=0.75\right)$. The code used to compute the kSZ power spectrum can be found at \url{https://github.com/adeliegorce/tools4reionisation}.

\section{Derivation of the kSZ angular power spectrum}
\label{sec:equations}

\subsection{Temperature fluctuations}

The CMB temperature anisotropies coming from the scattering of CMB photons off clouds of free electrons with a non-zero bulk velocity $\vvec$ relative to the CMB rest-frame along the line of sight $\los$ write
\begin{equation}
\label{eq:delta_TkSZ}
  \delta T_{k\rm SZ}(\los) =\frac{\sigma_T}{c} \int \frac{\drm \eta}{\drm z}\frac{\drm z}{(1+z)}\, \mathrm{e}^{-\tau(z)}\, n_e(z) \, \vvec\cdot\los \;,
\end{equation}
with $\sigma_T$ being the Thomson cross-section, $c$ the speed of light, $\eta$ the comoving distance to redshift $z$ and $\vvec\cdot\los$ the component of the peculiar velocity of the electrons along the line of sight. As mentioned before, $\tau$ is the Thomson optical depth, $\tau(z) = c\,\sigma_\mathrm{T} \int_{0}^{z} n_e(z')/H(z')\, (1+z')^2  \ \drm z'$. $n_e$ is the mean free electrons number density at redshift $z$ from which we derive the density contrast $\delta_e$ via $n_e = \bar{n}_e (1+\delta_e)$. We choose the limits of the integral in Eq~\eqref{eq:delta_TkSZ} depending on the type of signal we are interested in: for homogeneous kSZ, we integrate from $0$ to $z_{\rm end}$, the redshift when reionisation ends; for patchy kSZ, the main focus of this work, we integrate from $z_{\rm end}$ to the highest redshift considered in the simulation (here, $z_\mathrm{max}=15$). The contribution from redshifts larger than the onset of reionisation, when the only free electrons in the IGM are leftovers from recombination, is found to be negligible. 

We define $\qvecreal \equiv \vvec (1+\delta_e)  =\vvec + \vvec\delta_e \equiv \vvec + \qvecreal_e   $ the density-weighted peculiar velocity of the free electrons. It can be decomposed into a divergence-free $\qvecreal_B$ and a curl-free $\qvecreal_E$ components. We write their equivalents in the Fourier domain as $\qvec = \qvec_E + \qvec_B$. As pointed out by \citet{jaffe_1998}, when projected along the line of sight, $\qvec_E$ will cancel and only the component of $\qvec$ perpendicular to $\kvec$, that is $\qvec_B$, will contribute to the kSZ signal.
We want an expression for the kSZ angular power spectrum $C_\ell^{k \rm SZ} \equiv T^2_\mathrm{CMB} \vert \tilde{\delta T}_\mathrm{kSZ}(k)\vert^2$ where $k \equiv \ell / \eta$ is the Limber wave-vector and $\ell$ is the multipole moment, which can be related to an angular scale in the sky. In the small angle limit, the kSZ angular power spectrum can be derived from Eq.~\eqref{eq:delta_TkSZ} using the Limber approximation:
\begin{equation}
\begin{aligned}
\label{eq:def_C_ell_kSZ}
C_\ell =  & \frac{8 \pi^2}{(2\ell+1)^3} \frac{\sigma_T^2}{c^2} \int \frac{\bar{n}_e(z)^2}{(1+z)^2}\, \Delta_{B,e}^2(\ell/\eta,z)\, \exp^{-2 \tau(z)}\, \eta \, \frac{\drm \eta}{\drm z}\, \drm z ,
\end{aligned}
\end{equation}
with $\Delta_{B,e}^2(k,z)  \equiv k^3 P_{B,e}(k,z)/(2\pi^2)$ and $P_{B,e}$ the power spectrum of the curl component of the momentum field defined by $(2\pi)^3 P_{B,e}\, \delta_D(\kvec-\kvec')= \langle \qvec_{B,e}(\kvec)\ \qvec_{B,e}^*(\kvec')\rangle$ where $\delta_D$ is the Dirac delta function, the tilde denotes a Fourier transform and the asterisk a complex conjugate. 

Expanding $\langle \qvec_{B,e} \qvec_{B,e}^*\rangle$, we obtain:
\begin{equation}
  \qvec_{B,e}(\kvec) = \int \frac{\drm ^3 \kvec'}{(2\pi)^3} \, (\kvecunit' - \mu \kvecunit)\, \tilde{v}(k')\, \tilde{\delta}_e \left(|\kvec - \kvec'| \right),
\end{equation}
where $\mu = \kvecunit \cdot \kvecunit'$, so that
\begin{equation}
\label{eq:full_delta_B}
\begin{aligned}
  \frac{\langle \qvec_{B,e}(\kvec)\ \qvec_{B,e}^*(\kvec') \rangle}{(2\pi)^3{\delta_D}(|\kvec - \kvec'|)} \equiv & \frac{2\pi^2}{k^3} \Delta^2_{B,e}(k,z) \\
  = & \frac{1}{(2\pi)^3} \int \drm^3 k'\,  \left[ (1-\mu^2)\, \Pee (|\kvec-\kvec'|)\, P_{vv}(k') \right. \\ & \left. - \frac{(1-\mu^2)\, k'}{|\kvec-\kvec'|}P_{ev}(|\kvec-\kvec'|) \, P_{ev}(k') \, \right],
\end{aligned}
\end{equation}
where the $z$-dependencies have been omitted for simplicity. $\Pee(k,z)$ is the power spectrum of the free electrons density fluctuations and $P_{ev}$ is the free electrons density - velocity cross-spectrum.
In the linear regime, we can write $\vvec (\kvec) = i \kvec \, (f \dot{a} /k)\, \tilde{\delta}(\kvec)$, where $a$ is the scale factor and $f$ the linear growth rate defined by $f(a)= \drm \mathrm{ln}D / \drm \mathrm{ln}a$ for $D$ the growth function. With this we can compute the velocity power spectrum fully from linear theory and not be limited by the simulation size and resolution:
\begin{equation}
\label{eq:def_Pvv}
    P_{vv} (k,z) = \left( \frac{ \dot{a}f(z)}{k} \right) ^2 \Plin(k,z)
\end{equation}
where $\Pdd^\mathrm{lin}$ is the linear total matter power spectrum. We also assume for the cross-spectrum:
\begin{equation}
    P_{ve}(k,z) \simeq b_{\delta e}(k,z) P_{\delta v}(k,z) = \frac{f \dot{a}(z)}{k}\, b_{\delta e}(k,z) \Pdd^\mathrm{lin}(k,z),
\end{equation}
where the bias $b_{\delta e}$ is defined by the ratio of the free electrons power spectrum over the non-linear matter power spectrum $b_{\delta e} (k,z)^2 =\Pee(k,z) / \Pdd(k,z)$. Although coarse, this approximation only has a minor impact on our results: it implies variations of $\sim 0.05~\mu\mathrm{K}^2$ in the power spectrum amplitude \citep[see also][]{alvarez_2016}. The final expression of the power spectrum of the curl component of the momentum field then writes
\begin{equation}
\label{eq:delta_Be}
\begin{aligned}
  P_{B,e}(k,z) & = \frac{1}{(2\pi)^3} f(z)^2 \dot{a}(z)^2 \int \, \drm^3 k' (1-\mu^2)  \times \\ & \left[ \, \frac{ 1}{k'^2} \Pee (|\kvec-\kvec'|)\,  \Pdd^\mathrm{lin}(k',z) \right. 
  \\ & \left. - \frac{b_{\delta e}(k',z)}{|\kvec-\kvec'|^2}\,b_{\delta e}(|\kvec-\kvec'|,z) \,  P_{\delta \delta}^\mathrm{lin} (|\kvec-\kvec'|,z) \, P_{\delta \delta}^\mathrm{lin}(k',z) \right],
\end{aligned}
\end{equation}
which we can plug into Eq.~\eqref{eq:def_C_ell_kSZ} to find the final expression for the kSZ angular power spectrum.

\subsection{The power spectrum of free electrons density contrast}

\begin{figure}
    \centering
    \includegraphics[width=0.9\columnwidth]{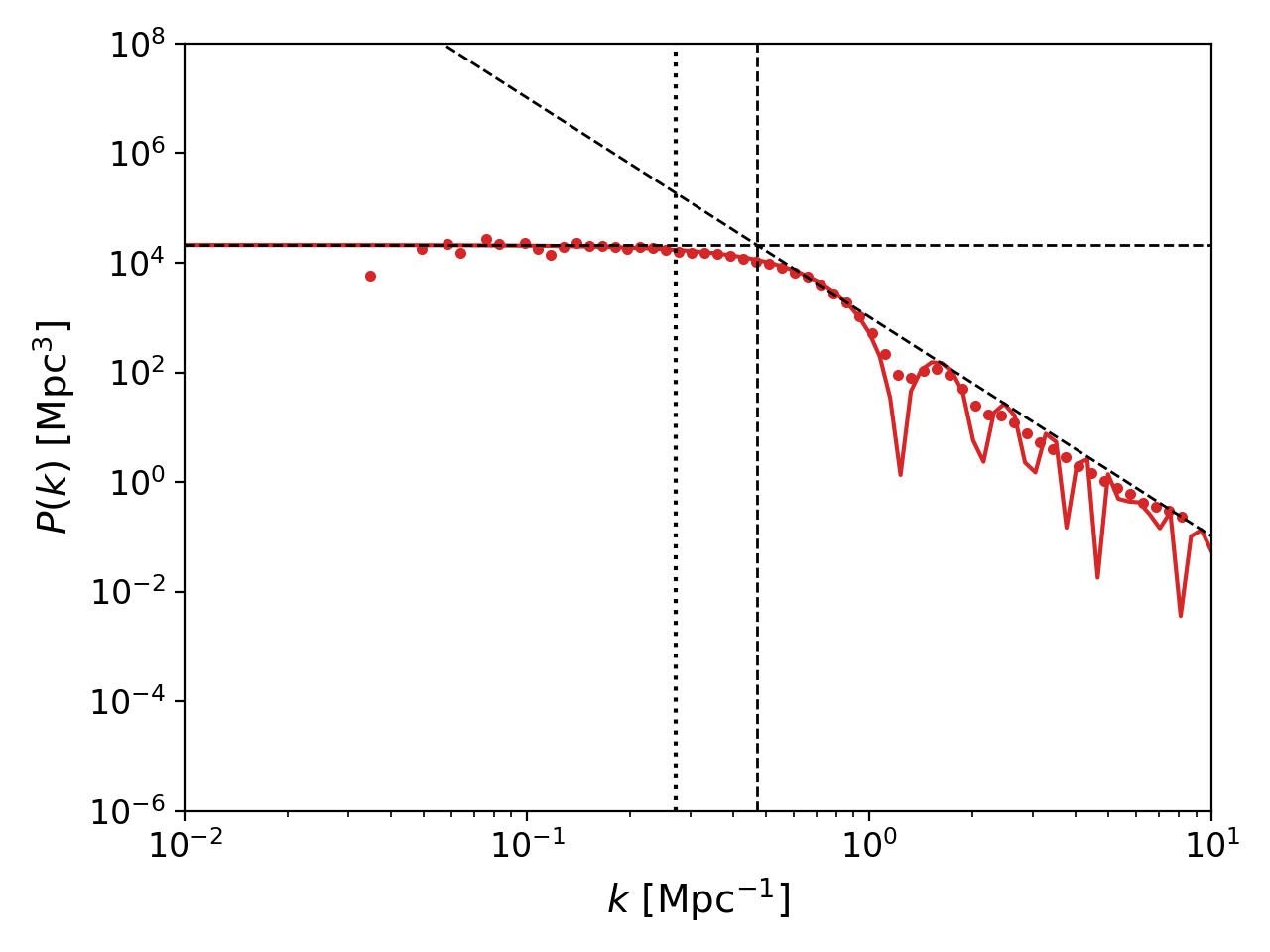}
    \caption{Free electrons density contrast power spectrum for a box filled with enough bubbles of radius $R = 15~\mathrm{px} = 5.5~\mathrm{Mpc}$ to reach a filling fraction $f= 1 \%$. Points are results of a numerical computation of the power spectrum, compared to the theoretical model (solid line). The dotted vertical line corresponds to $k=1/R$, the dashed vertical line to $9^{1/4}/R$, the dashed horizontal line to $4/3 \pi R^3 / f$ and the tilted dashed line has slope $k^{-4}$.}
    \label{fig:PS_bubbles}
\end{figure}

In \citet{shaw_2012}, the authors choose to describe the behaviour of the free electrons power spectrum in terms of a biased matter power spectrum: they take $\Pee (k,z) \equiv b_{\delta e}(k,z)^2 \Pdd (k,z)$ and calibrate $b_{\delta e}(k,z)$ on their simulations, either extrapolating or assuming a reasonable behaviour for the scales and redshifts not covered by the simulations. However, because $\Pee$ describes the free electrons density fluctuations, it has a relatively simple structure, close to the power spectrum of a field made of ionised spheres on a neutral background, shown in Fig.~\ref{fig:PS_bubbles}, and using a bias is not necessary.

\begin{figure*}
    \centering
    \includegraphics[height=5cm]{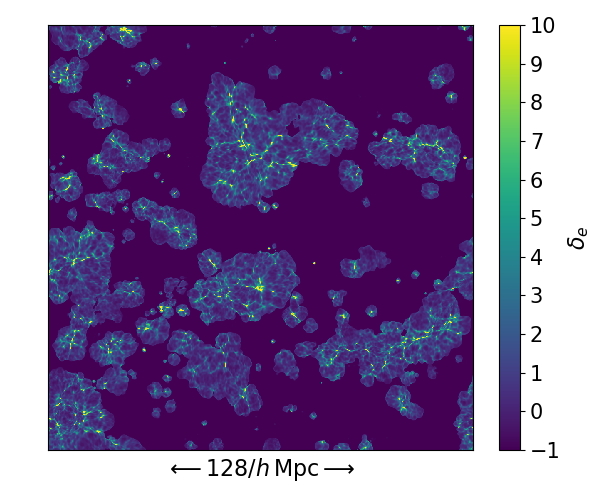} 
    \includegraphics[height=5cm]{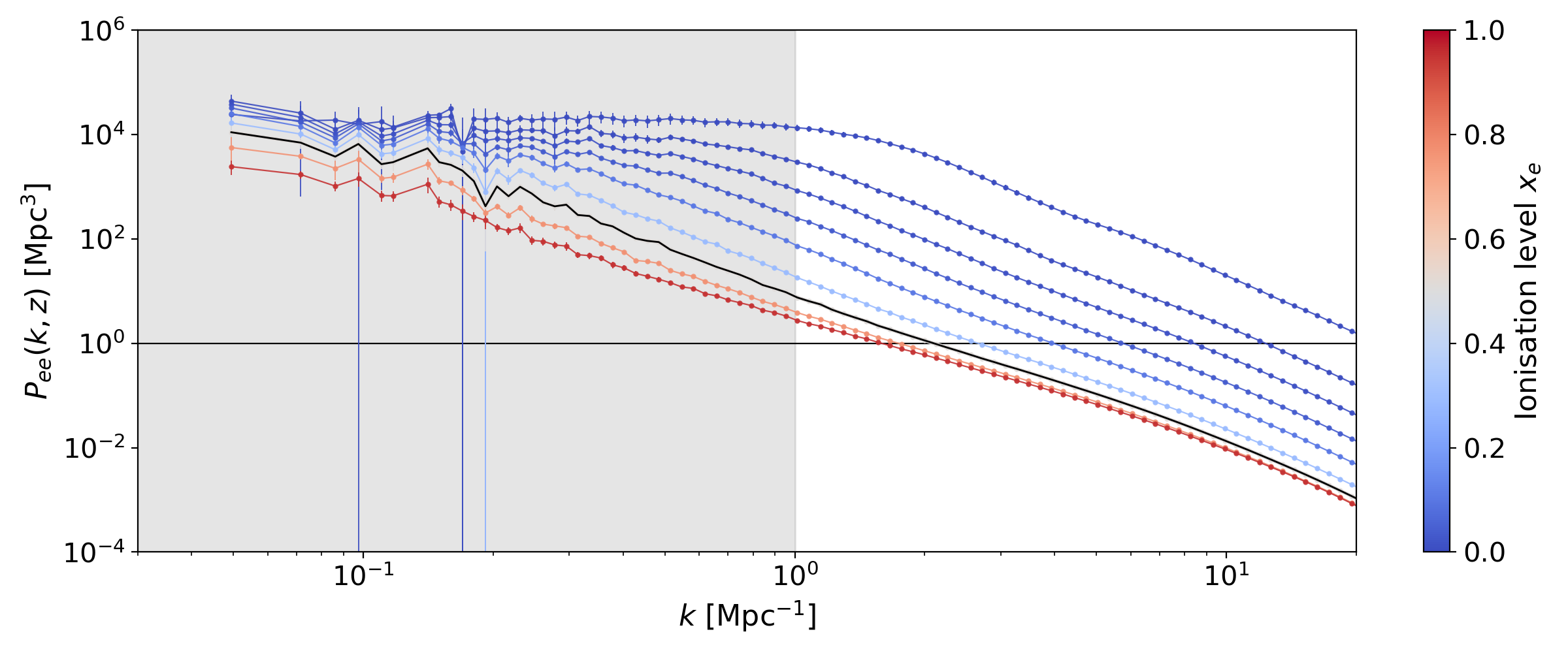}
    \caption{\textit{Left panel:} Snapshot of the electron density contrast field for the first of the six simulations, at $z=7.2$ and $x_e = 0.49$. \textit{Right panel:} Free electrons power spectrum of the same simulation at fixed redshifts (fixed ionised levels). The shaded area corresponds to scales contributing $\dthree^\mathrm{patchy}$ the most (see Sec.~\ref{subsec:fit}) and the solid black line to the field shown in the left panel.}
    \label{fig:electron_field_sim1}
\end{figure*}

Consider a box of volume $V=L^3$ filled with $n$ fully ionised bubbles of radius $R$, randomly distributed throughout the box so that their centres are located at $ \bm{a} _i$ for $i \in \{ 1, n \}$. 
The density of free electrons in the box follows
\begin{equation}
n_e(\bm{r}) = \frac{\bar{n}_e}{f} \sum_{i=1}^n \Theta \left( \frac{ \vert \bm{r} - \bm{a}_i \vert}{R} \right),
\end{equation}
where $\Theta \left( x\right)$ is the Heaviside step function, $\bar{n}_e$ is the mean number density of electrons in the box and $f$ the filling fraction of the box (here, $f = \xhii$). $\bar{n}_e/f$ is the number of electrons in one bubble divided by its volume and, ignoring overlaps, $f = 4/3 \pi R^3 n / V$.  Consider the electron density contrast field $\delta_e$ on which $\Pee(k,z)$ is built:
\begin{equation}
    \delta_e (\bm{r}) = \frac{n_e(\bm{r})}{\bar{n}_e} - 1 = \frac{1}{f} \sum_{i=1}^n \Theta \left( \frac{ \vert \bm{r} - \bm{a}_i \vert}{R} \right) - 1,
\end{equation}
represented on Fig.~\ref{fig:electron_field_sim1} for one of the simulations used in this work. $\delta_e(\bm{r})$ Fourier--transforms into
\begin{equation}
\tilde{\delta}_e ( \bm{k} ) = \frac{L^3}{n}\, W(kR)  \sum_{i=1}^n \mathrm{e}^{- i \bm{k} \cdot \bm{a}_i},
\end{equation}
where $W$ is the spherical top-hat window function $W(y) = (3/y^3)\left[ \sin y - y\, \cos y\right]$. Using this expression, and following \citet{bharadwaj_2005}, the power spectrum of the electron density contrast field writes:
\begin{equation}
\label{eq:PS_toy_model}
\Pee(\bm{k}) =  \frac{4}{3}\pi R^3 \, \frac{1}{f} W^2(kR),
\end{equation}
which has units $\mathrm{Mpc}^3$. Fig.~\ref{fig:PS_bubbles} gives an example of such a power spectrum. We have generated an ionisation field made of enough bubbles of radius $R = 15~\mathrm{px} = 5.5~\mathrm{Mpc}$\footnote{The bubble radii actually follow a Gaussian distribution centred on $15~\mathrm{px}$ with standard deviation $2~\mathrm{px}$.} to reach a filling fraction $f= 1 \%$ in a box of $512^3$ pixels and side length $L = 128/h ~\mathrm{Mpc}$. We compare the expression in Eq.~\eqref{eq:PS_toy_model} with power spectrum values computed directly from the 3D field and find a good match. On very small or very large scales, the window function behaves as:
\begin{equation}
\begin{aligned}
& W(y) \sim \frac{3}{y^3} \times \frac{y^3}{3} =1 &&\mathrm{as}\ y \rightarrow 0 \\
& W(y) \sim \frac{3}{y^3} \times y = \frac{3}{y^2} &&\mathrm{as}\ y \rightarrow \infty
\end{aligned}
\end{equation}
so that $\Pee(k) \sim  4/3 \pi R^3 / f$ is constant (see dashed horizontal line on the figure) on very large scales and has higher amplitude for smaller filling fractions. On small scales, the toy model power spectrum decreases as $k ^{-4}$ (see tilted dashed line on the figure). The intersection point of the horizontal and tilted dashed lines on the figure corresponds to $k = 9^{1/4}/R$ (dashed vertical line), hinting at a relation between the cut-off frequency and the bubble size. Interestingly, \citet{xu_2019_HI_bias} find a similar feature, also related to the typical bubble size, in the bias between the $\ion{H}{I}$ and matter fields. 

This behaviour is close to what we observe in the free electrons density power spectra of the custom set of simulations used in this work in the early stages of reionisation, as can be seen on the right panel of Fig.~\ref{fig:electron_field_sim1}. Therefore, we choose in this work to use a direct parameterisation of the scale  and redshift evolution of $\Pee(k,z)$ during reionisation and calibrate it on our simulations. The parameters, $\alpha_0$ and $\kappa$, are defined according to:
\begin{equation}
\label{eq:fit_formula}
    \Pee(k,z) = \frac{\alpha_0 \ x_e(z)^{-1/5}}{1 + [k/ \kappa]^{3} x_e(z)}.
\end{equation}
In log-space, on large scales, $\Pee$ has a constant amplitude which, as mentioned above, depends on the filling fraction and therefore reaches its maximum $\alpha_0$ at the start of the reionisation process, when the variance in the free electron field is maximal (see Sec.~\ref{subsec:physical_interpretation}). It then slowly decreases as $x_e(z)^{-1/5}$. Before the onset of reionisation, despite the few free electrons left over after recombination, the amplitude of $\Pee$ is negligible. This constant power decreases above a cut-off frequency that increases with time, following the growth of ionised bubbles, according to $\kappa x_e(z)^{-1/3}$. There is no power above this frequency, that is on smaller scales: there is no smaller ionised region than $r_\mathrm{min}(z)=2\pi x_e^{1/3} /\kappa$ at this time. For empirical reasons, we choose the power to decrease as $k^{-3}$ and not $k^{-4}$ as seen in the theoretical power spectrum on small scales. This difference can be explained by the fact that in our simulations, small ionised regions will keep appearing as new sources light up, maintaining power on scales smaller than the typical bubble size. Additionally, the density resolution will allow correlations between regions within a given bubble, whereas in the toy models ionised bubbles are only filled with ones. The complexity of the electron density contrast field is illustrated for one of the six simulations used in this work on Fig.~\ref{fig:electron_field_sim1}: the underlying matter field is visible within the ionised regions.

Once reionisation is over and all IGM atoms are ionised, the fluctuations in free electrons density follow those of dark matter on large scales ($k < 1 ~\mathrm{Mpc}^{-1}$). On smaller scales, gas thermal pressure induces a drop in $\Pee(k,z)$ compared to the dark matter. To describe this evolution at low redshifts, we choose the same parameterisation as \citet{shaw_2012}, given in Eq.~\eqref{eq:bias_shaw}, to describe the gas bias $b_{\delta e}(k,z)^2 = \Pee (k,z) / \Pdd (k,z)$ but adapt the parameters to our simulations, which however do not cover redshifts lower than $5.5$:
\begin{equation}
\label{eq:bias_shaw}
    b_{\delta e}(k,z)^2 = \frac{1}{2} \left[ \exp^{-k/k_f} + \frac{1}{1 + (g k/k_f)^{7/2} } \right].
\end{equation}
We find $k_f = 9.4 ~\mathrm{Mpc}^{-1}$ and $g = 0.5$, constant with redshift. Our values for $k_f$ and $g$ are quite different from those obtained by \citet{shaw_2012}, as in their work power starts dropping between $0.05$ and $0.5 ~\mathrm{Mpc}^{-1}$ compared to $k \sim 3 ~\mathrm{Mpc}^{-1}$ for our simulations. This can be explained by our simulations making use of adaptive mesh refinement, therefore resolving very well the densest regions, so that our spectra are more sensitive to the thermal behaviour of gas. This model, where $k_f$ and $g$ are constant parameters, is a very basic one. It will however be sufficient for this work since we focus on the patchy component of the kSZ effect, at $z \geq 5.5$. Additionally, as shown later, the scales mostly contributing to the patchy kSZ signal correspond to modes $10^{-3} < k / \mathrm{Mpc}^{-1} < 1$ where $\Pee$ follows the matter power spectrum, so that a precise knowledge of $b_{\delta e}(k,z)$ is not required. In the future, if we want to apply our results to constrain reionisation with the measured CMB temperature power spectrum, we will need a better model as the observed signal will be the sum of homogeneous and patchy kSZ, with the former dominating on all scales.

To account for the smooth transition of $\Pee$ from a power-law to a biased matter power spectrum, illustrated in the right panel of Fig.~\ref{fig:electron_field_sim1}, we write the final form for the free electrons density fluctuations power spectrum as 
\begin{equation}
\label{eq:full_Pee_expression}
\begin{aligned}
    \Pee (k,z) = \left[ f_\mathrm{H} - x_e(z) \right] &  \times \frac{\alpha_0 \ x_e(z)^{-1/5}}{1 + [k/ \kappa]^{3} x_e(z)} \\ & + \ x_e(z) \times b_{\delta e}(k,z)^2 \Pdd (k,z),
\end{aligned}
\end{equation}
for $f_\mathrm{H} = 1+ Y_p/4X_p \simeq 1.08$, with $Y_p$ and $X_p$ the primordial mass fraction of helium and hydrogen respectively. 
The total matter power spectrum $\Pdd$ is computed using the Boltzmann integrator \texttt{CAMB} \citep{camb1,camb2} for the linear terms and the \texttt{HALOFIT} procedure for the non-linear contributions \citep{smith_2003_halofit}.

\section{Calibration on simulations}
\label{sec:calibration}

\subsection{Description of the simulations}
\label{subsec:simulations}

The simulations we use in this work were produced with the EMMA simulation code \citep{aubert_2015_EMMA} and previously used in \citet{chardin_2019}. The code tracks the collisionless dynamics of dark matter, the hydrodynamics of baryons, star formation and feedback, and the radiative transfer using a moment-based method \citep[see][]{aubert_2018,deparis_2019}. This code adheres to an Eulerian description, with fields described on grids, and enables adaptive mesh refinement techniques to increase the resolution in collapsing regions. Six simulations with identical numerical and physical parameters were produced in order to make up for the limited physical size of the box and the associated sample variance. They only differ in the random seeds used to generate the initial displacement phases, resulting in 6 different configurations of structures within the simulated volumes. Each run has a $(128~\mathrm{Mpc}/h)^3$ volume sampled with $1024^3$ cells at the coarsest level and $1024^3$ dark matter particles. Refinement is triggered when the number of dark matter particles exceeds 8, up to 6 refinement levels. Initial conditions were produced using MUSIC \citep{2013ascl.soft11011H} with a starting redshift of $z=150$, assuming \citet{planck_2015_overview} cosmology. Simulations were stopped at $z\sim6$, before the full end of reionisation. The dark matter mass resolution is $2.1 \times 10^8 \mathrm{M}_\odot$ and the stellar mass resolution is $6.1 \times 10^5 \mathrm{M}_\odot$. Star formation proceeds according to standard recipes described in \citet{rasera_2006}, with an overdensity threshold equal to 20 to trigger the gas-to-stellar particle conversion with a 0.1 efficiency: such values allow the first stellar particles to appear at $z \sim 17$. Star particles produce ionising radiation for 3 Myr, with an emissivity provided by the Starburst99 model for a Top-Heavy initial mass function and a $Z=0.001$ metallicity \citep{starbust99}. Supernova feedback follows the prescription used in \citet{aubert_2018}: as they reach an age of 15 million years, stellar particles dump $9.8 \times 10^{11}~\mathrm{J}$ per stellar kg in the surrounding gas, 1/3 in the form of thermal energy, 2/3 in the form of kinetic energy. Using these parameters, we obtain a cosmic star formation history consistent with constraints by \citet{bouwens_2015} and end up with 20 millions stellar particles at $z=6$. The simulations were produced on the Occigen (CINES) and Jean-Zay (IDRIS) supercomputers, using CPU architectures : a reduced speed of light of $0.1c$ has been used to reduce the cost of radiative transfer. 

\begin{table}
    \caption{Characteristics of the six high resolution simulations used. $\zrei$ is the midpoint of reionisation $\xhii(\zrei) = 0.5 f_\mathrm{H}$, $z_\mathrm{end}$ the redshift at which $x_e(z)$ (extrapolated) reaches $f_\mathrm{H}$ and $\tau$ is the Thompson optical depth. $\Delta z$ corresponds to $z_{0.25} - z_{0.75}$.}
    \label{tab:six_simulations}
    \centering
    \begin{tabular}{c|ccccccccc}
            & $z_\mathrm{re}$ & $z_\mathrm{end}$ & $\tau_{\xhii}$ & $\Delta z$\\
        \hline 
        1 & 7.09 & 5.96 & 0.0539 & 1.17 \\
        2 & 7.16 & 5.92 & 0.0545 & 1.19 \\
        3 & 7.16 & 5.67 & 0.0544 & 1.16 \\
        4 & 7.05 & 5.60 & 0.0532 & 1.16 \\
        5 & 7.03 & 5.56 & 0.0531 & 1.15 \\
        6 & 7.14 & 5.79 & 0.0543 & 1.16 \\
        Mean & 7.10 & 5.84 & 0.0541 & 1.16
    \end{tabular}
\end{table}

\begin{figure*}[!h]
    \centering
    \includegraphics[height=4.5cm]{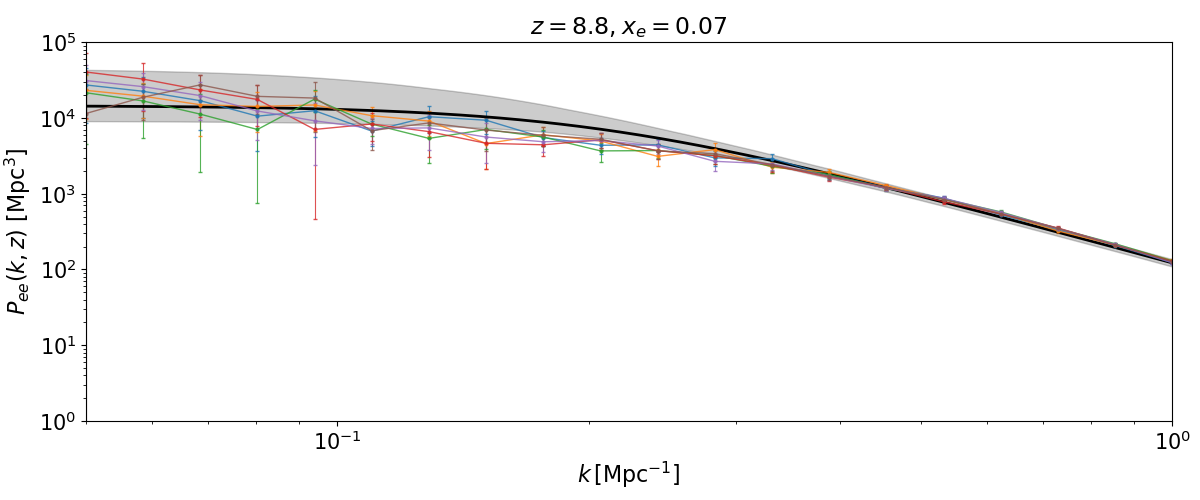}
    \includegraphics[height=4.5cm]{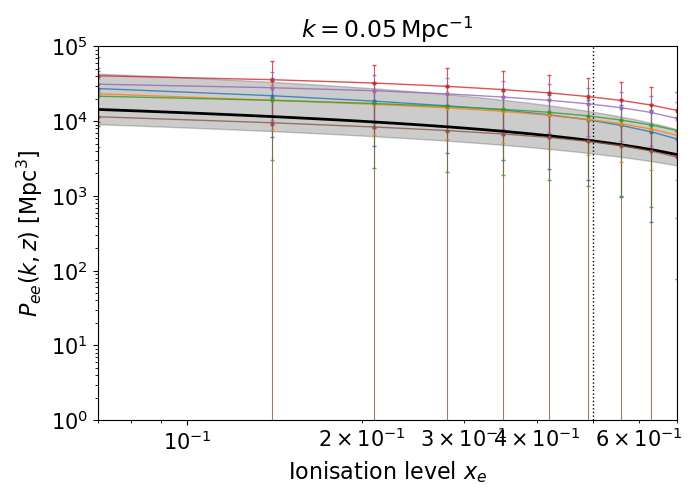} \\
    \includegraphics[height=4.5cm]{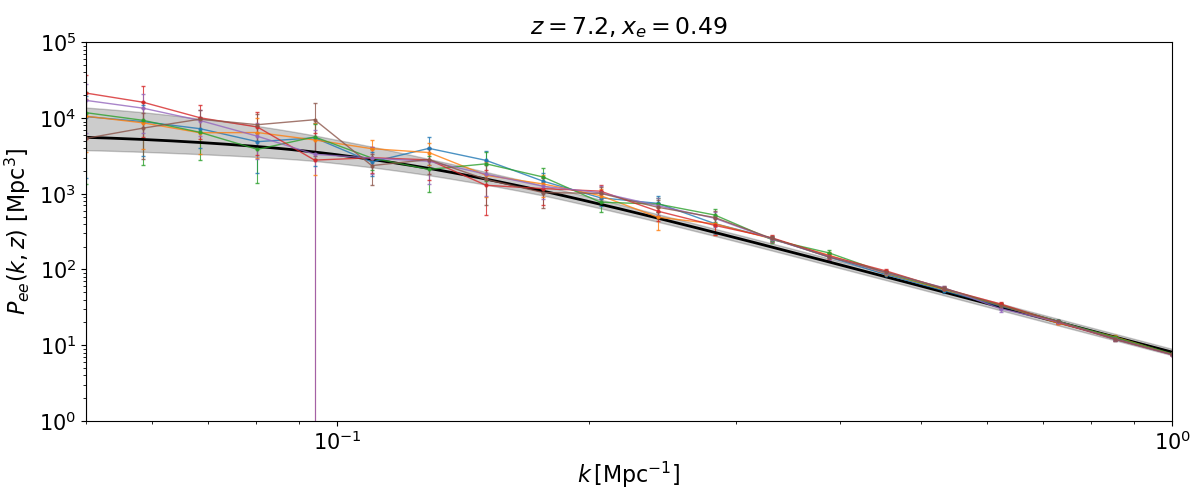} 
    \includegraphics[height=4.5cm]{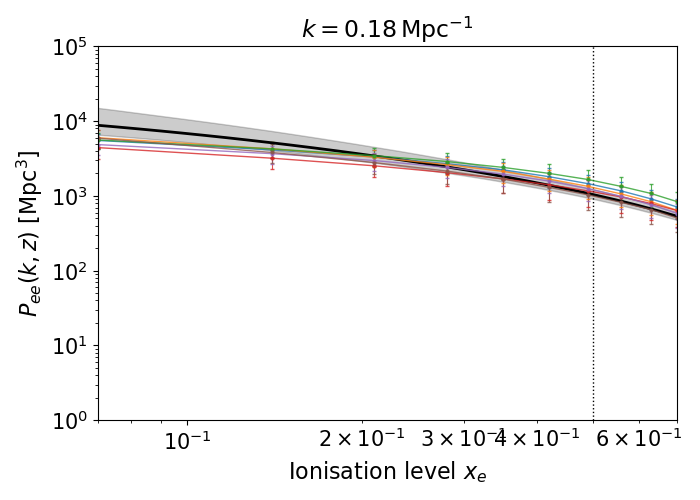}\\
    \includegraphics[height=4.5cm]{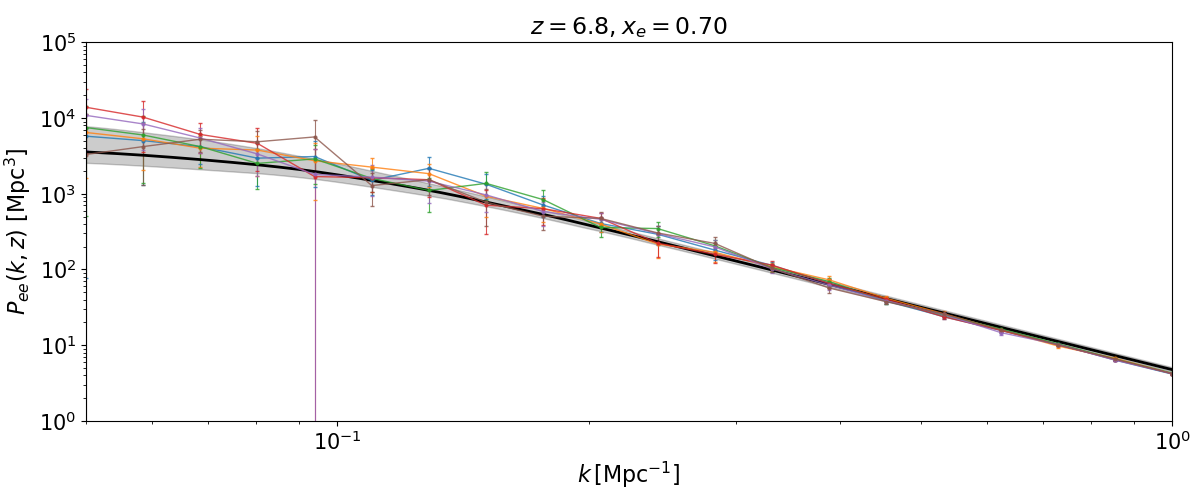}
    \includegraphics[height=4.5cm]{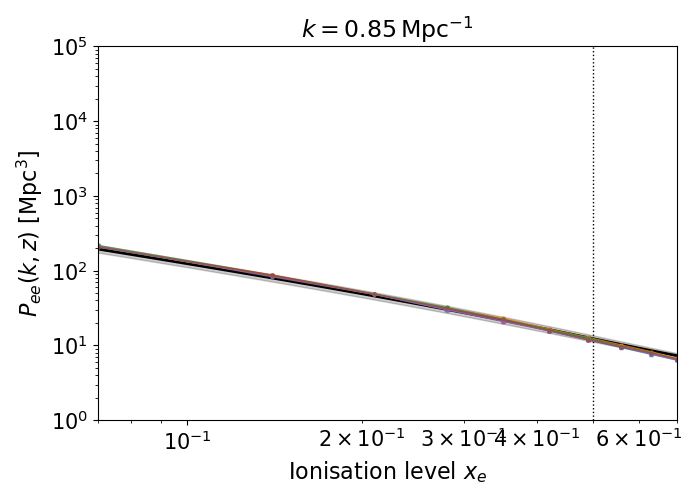}
    \caption{Result of the fit of Eq.~\eqref{eq:full_Pee_expression} on the free electrons power spectrum of our six simulations, for three redshift bins (left panels) and three scale bins (right panels). The best-fit is shown as the thick black line with the accompanying $68\%$ confidence interval, and the spectra of the six simulations as thin coloured lines. Error bars on data points are computed from the covariance matrix (see text for details).}
    \label{fig:Pee_vs_k}
\end{figure*}

Table \ref{tab:six_simulations} gives the midpoint $\zrei$ and end of reionisation $z_\mathrm{end}$ for each simulation, as well as the duration of the process, defined as the time elapsed between  global ionisation fractions of $25\%$ and of $75\%$\footnote{Some of our simulations end before reionisation is achieved, therefore we extrapolate $x_e(z)$ to find the $\zend$ value.}. The upper panel of Fig.~\ref{fig:ksz_six_sims} shows the interpolated reionisation histories, where data points correspond to the snapshots available for each simulation. Originally, our simulations do not include the first reionisation of helium. We correct for this by multiplying the IGM ionised fraction of hydrogen $x_\ion{H}{II}$ measured in the simulations by $f_\mathrm{H} = 1+ Y_p/4X_p \simeq 1.08$. Because we limit our study to redshifts $z > 5.5$, the second reionisation of helium is ignored. Fig.~\ref{fig:electron_field_sim1} shows the electron density contrast field for the first of our six simulations, close to the midpoint of reionisation. The complexity of the structure of this field is summarised in its power spectrum, shown in the right panel. Fig.~\ref{fig:Pee_vs_k} compares the $\Pee(k,z)$ spectra of the six simulations, taken either at fixed redshift (first column) or fixed scale (right column). Despite identical numerical and physical parameters and very similar reionisation histories, the six simulations have different free electrons density power spectra, which translates into different kSZ power spectra.

\subsection{Calibration procedure}
\label{subsec:fit}

We simultaneously fit the power spectra of the six simulations to Eq.~\eqref{eq:full_Pee_expression} on a scale range $0.05 < k/\mathrm{Mpc}^{-1} < 1.00$ (20 bins), corresponding to the scales which contribute the most to the signal at $\ell = 3000$ (see next paragraph), and a redshift range of $6.5 \leq z \leq 10.0$ (10 bins), corresponding to the core of the reionisation process ($0.07 < \xhii < 0.70$).\footnote{Because the snapshots of each simulation are not taken at the same redshifts or ionisation levels, we interpolate $\Pee(k,z)$ for each simulation and then compute the interpolated spectra for a common set of ionisation levels, with less elements than the original number of snapshots. Note that the original binning in scales for $\Pee(k,z)$ is the same for the six simulations but reduced from 38 to 20 bins.} We sample the parameter space of $\alpha_0$ and $\kappa$ on a regular grid (with spacings $\Delta \log \alpha_0 = 0.001 $ and $\Delta \kappa = 0.0001$) for which we compute the following likelihood:
\begin{equation}
    \chi^2 = \sum_\mathrm{n = 1}^{6} \sum_{z_i} \sum_{k_j} \frac{1}{\sigma_e^2} \left[ \Pee^\mathrm{data}(k_j,z_i) - \Pee^\mathrm{model}(k_j,z_i) \right]^2 ,
\end{equation}
where $\left\{z_i\right\}$ and $\{k_j \}$ are the redshift and scale bins and the first sum is over the six simulations. Because our sample of six simulations is not sufficient to derive a meaningful covariance matrix, we choose to ignore correlations between scales across redshifts and use the diagonal of the covariance matrix to derive error bars $\sigma_e$ for each data point. We refer the interested reader to a discussion of this choice in Appendix \ref{app:cov_matrices}. We choose the best-fit as the duplet $(\alpha_0,\kappa)$ for which the reduced $\chi^2$ reaches its minimum value of $1.05$\footnote{The raw value is $\chi^2 \sim 2500$.}. The best-fit values, with their $68\%$ confidence intervals are
\begin{equation}
\begin{aligned}
& \log \alpha_0 /\mathrm{Mpc}^3 = 3.93 ^{+0.05}_{-0.06} \\
& \kappa = 0.084 ^{+0.003}_{-0.004}\, \mathrm{Mpc}^{-1}. \\
\end{aligned}
\end{equation}
We note a strong correlation between the two parameters due to both physical -- see Sec.~\ref{subsec:physical_interpretation} -- and analytical reasons. Indeed, the value of $\kappa$ impacts the low-frequency amplitude of the $\Pee(k,z)$ model. The best-fit model, compared to the $\Pee(k,z)$ spectra of the six simulations Eq.~\eqref{eq:full_Pee_expression} is fitted on, can be seen in Fig.~\ref{fig:Pee_vs_k} for three different redshift bins (left-hand column) and three different scale bins (right-hand column). Overall, we see a good agreement between the fit and the data points on the scales of interest, despite the simplicity of our model. 

\begin{figure*}
    \centering
    \includegraphics[height=5cm]{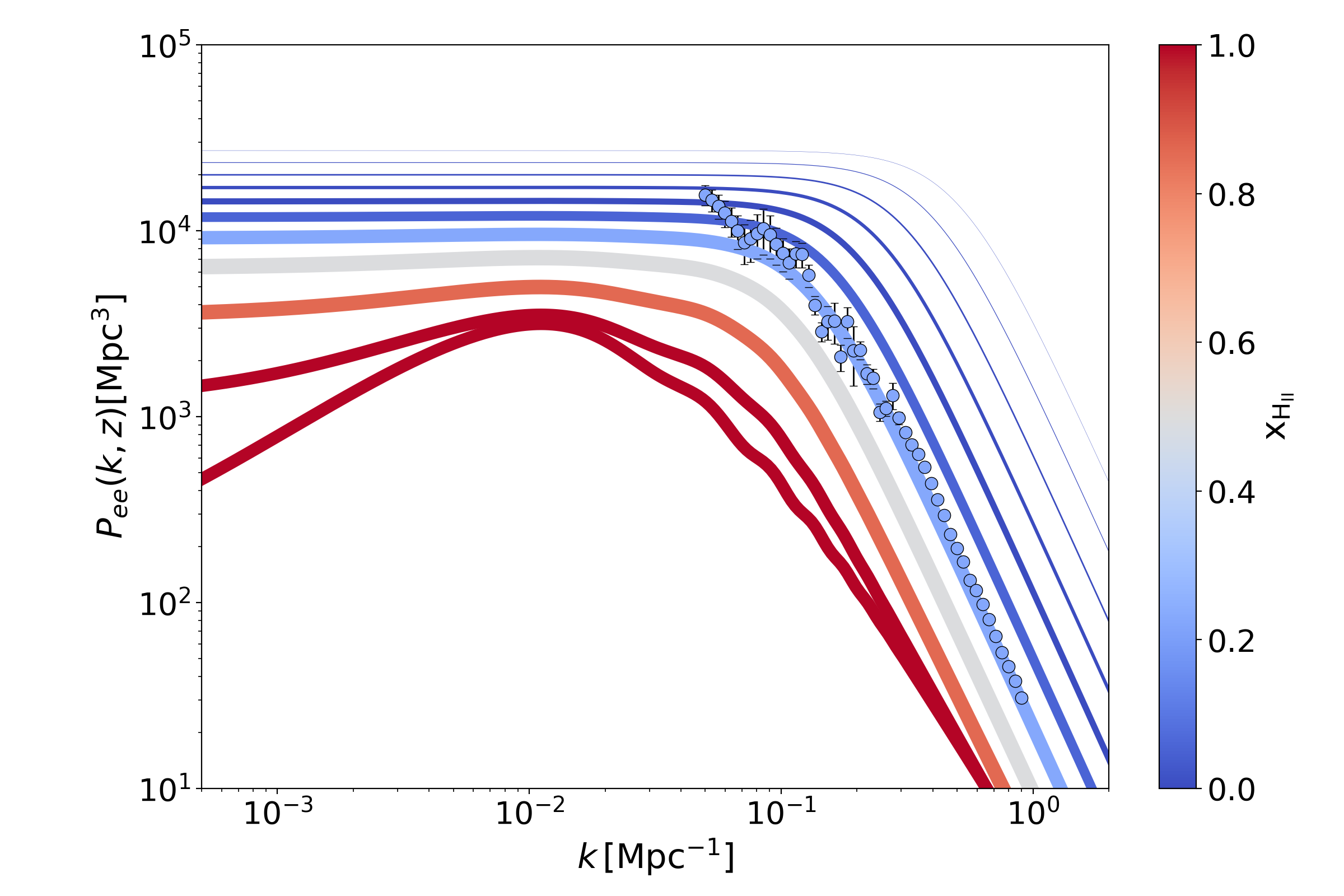}
    \includegraphics[height=5cm]{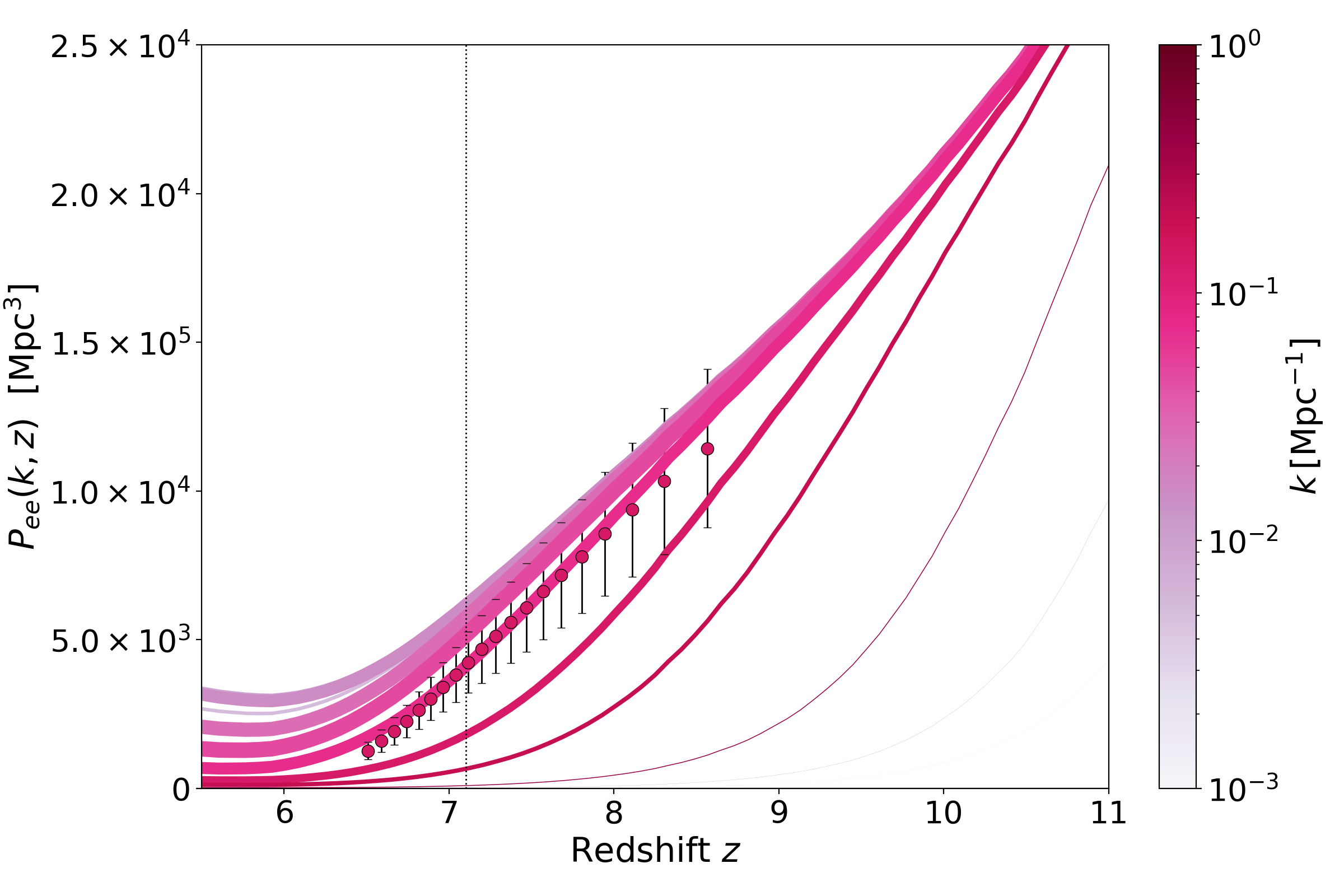}\\
    \includegraphics[height=5cm]{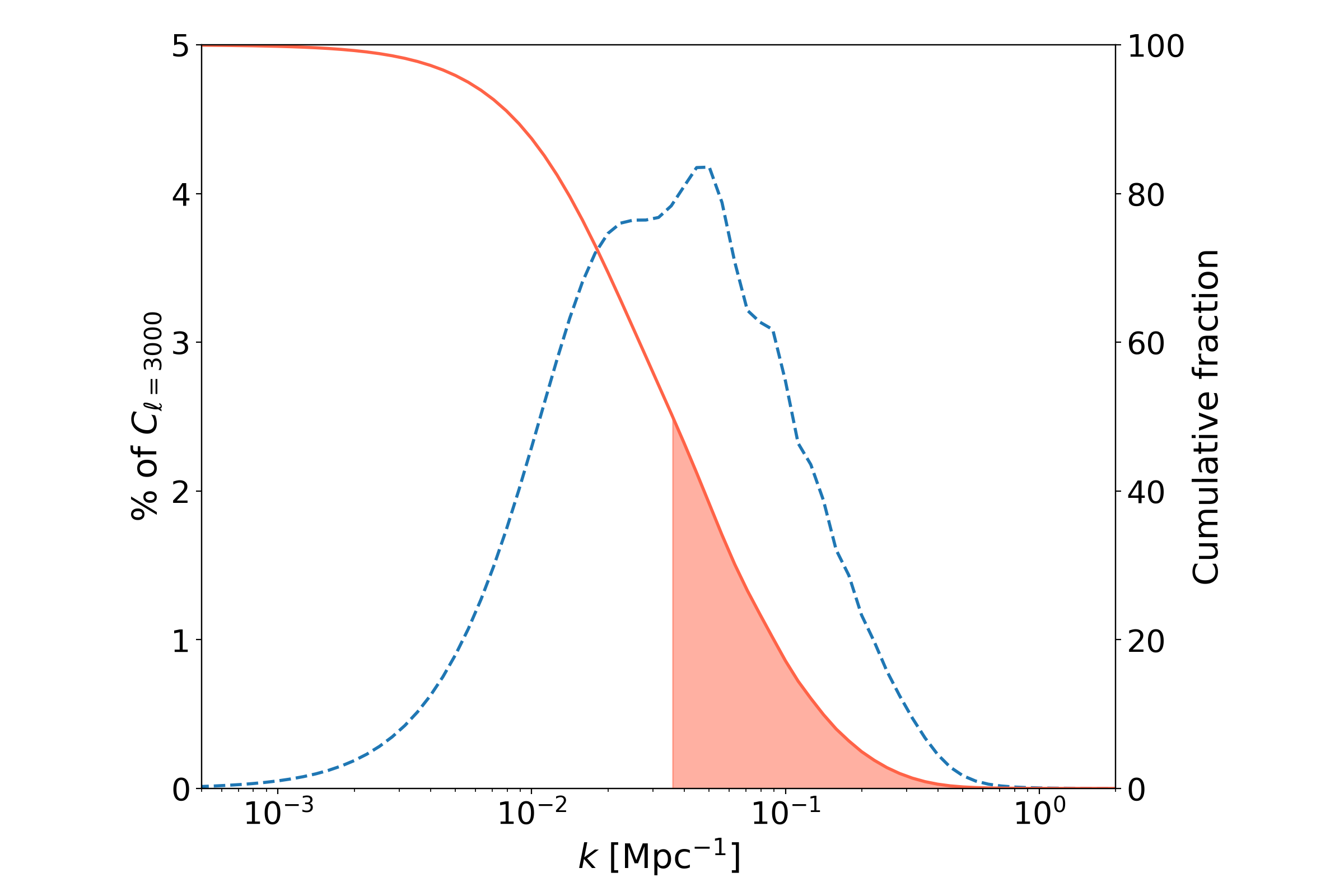}
    \includegraphics[height=5cm]{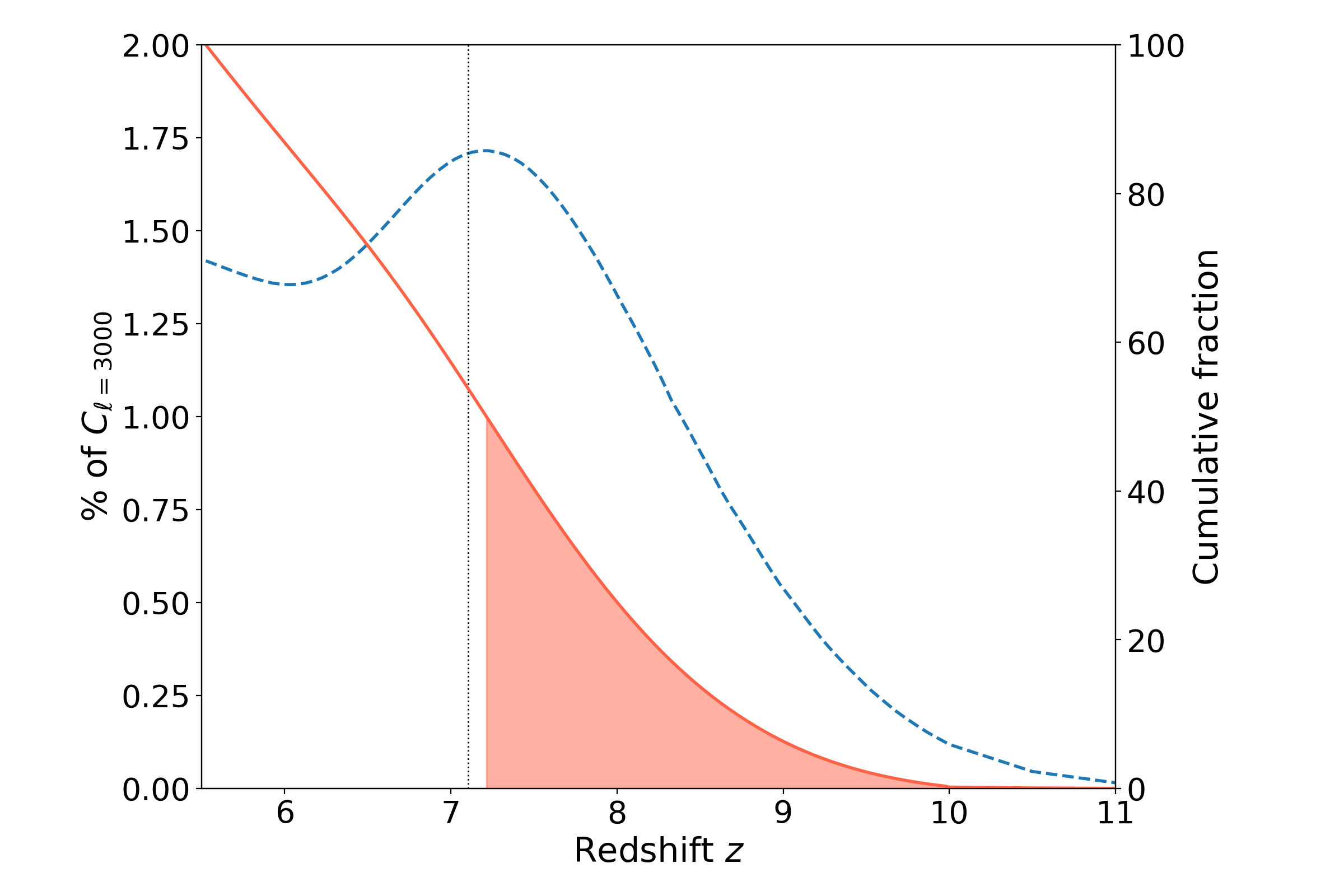}
    \caption{\textit{Upper panels:} $\Pee(k,z)$ as fitted on spectra from the fourth simulation, as a function of scales (left panel) and of redshift (right panel). For reference, the fit is compared to data points for $z=7.8$ ($x_e=0.26$) and $k=0.14~\mathrm{Mpc}^{-1}$ respectively, with corresponding colour. The width of each line represents the contribution of the redshift (resp. scale) of the corresponding scale (resp. redshift) to the final patchy kSZ amplitude at $\ell=3000$. \textit{Lower panels:} Corresponding probability densities (dashed lines) and cumulative distributions (solid lines). Shaded areas correspond to the first $50\%$ of the signal. The dotted vertical line on the lower right panel marks the midpoint of reionisation.}
    \label{fig:contributions}
\end{figure*}

Given the large number of $\Pee(k,z)$ data points originally ($\sim 3500$) and the complexity of the evolution of $\Pee$ with $k$ and $z$, we must limit our fits to given ranges. In order to assess what scales and redshifts contribute the most to the final kSZ signal, we look at the evolution of the integrand on $z$ in Eq.~\eqref{eq:def_C_ell_kSZ} with time and at the evolution of the integral on $k$ in Eq.~\eqref{eq:delta_Be} with scales. The results are shown in Fig.~\ref{fig:contributions}. 
The left (resp. right) upper panel presents the evolution of $\Pee(k,z)$ with scales (resp. redshift) after applying the fitting procedure described above. The width of each line represents the contribution of the redshift (resp. scale) of the corresponding colour to the final patchy kSZ amplitude at $\ell=3000$. The lower panels present the corresponding probability density and cumulative distribution functions. We find that redshifts throughout reionisation contribute homogeneously to the signal, since $50\%$ stems from redshifts $z \leq 7.2$, slightly before the midpoint $\zrei = 7.0$. Redshifts on the range $6.5<z<8.5$ contribute the most as they represent about $75\%$ of the final kSZ power. Conversely, redshifts $z>10$ contribute to only $0.4\%$ of the total signal. 
On the lower panel, we see that scales outside the range $10^{-3}~\mathrm{Mpc}^{-1} < k < 1~\mathrm{Mpc}^{-1}$ contribute very marginally to the final signal (about $~0.2\%$), whereas the range $10^{-2}< k/\mathrm{Mpc}^{-1} < 10^{-1}$ makes up about $70\%$ of $\dthree$.
Therefore, we choose to only keep data within the redshift range $6.5 < z< 10.0$ (i.e. $7\% < x_e < 70\%$) and the scale range $10^{-3}< k/\mathrm{Mpc}^{-1} < 1$ to constrain our fits.
For reference, on Fig.~\ref{fig:contributions}, we compare the fit to data points at $z=7.8$ ($x_e=0.26$) and $k=0.14~\mathrm{Mpc}^{-1}$ for the first simulation, and find an overall good match.

\section{Propagation to the kSZ power spectrum}
\label{sec:results}

\subsection{Results on our six simulations}

\begin{figure}
    \centering
    \includegraphics[width=0.8\columnwidth]{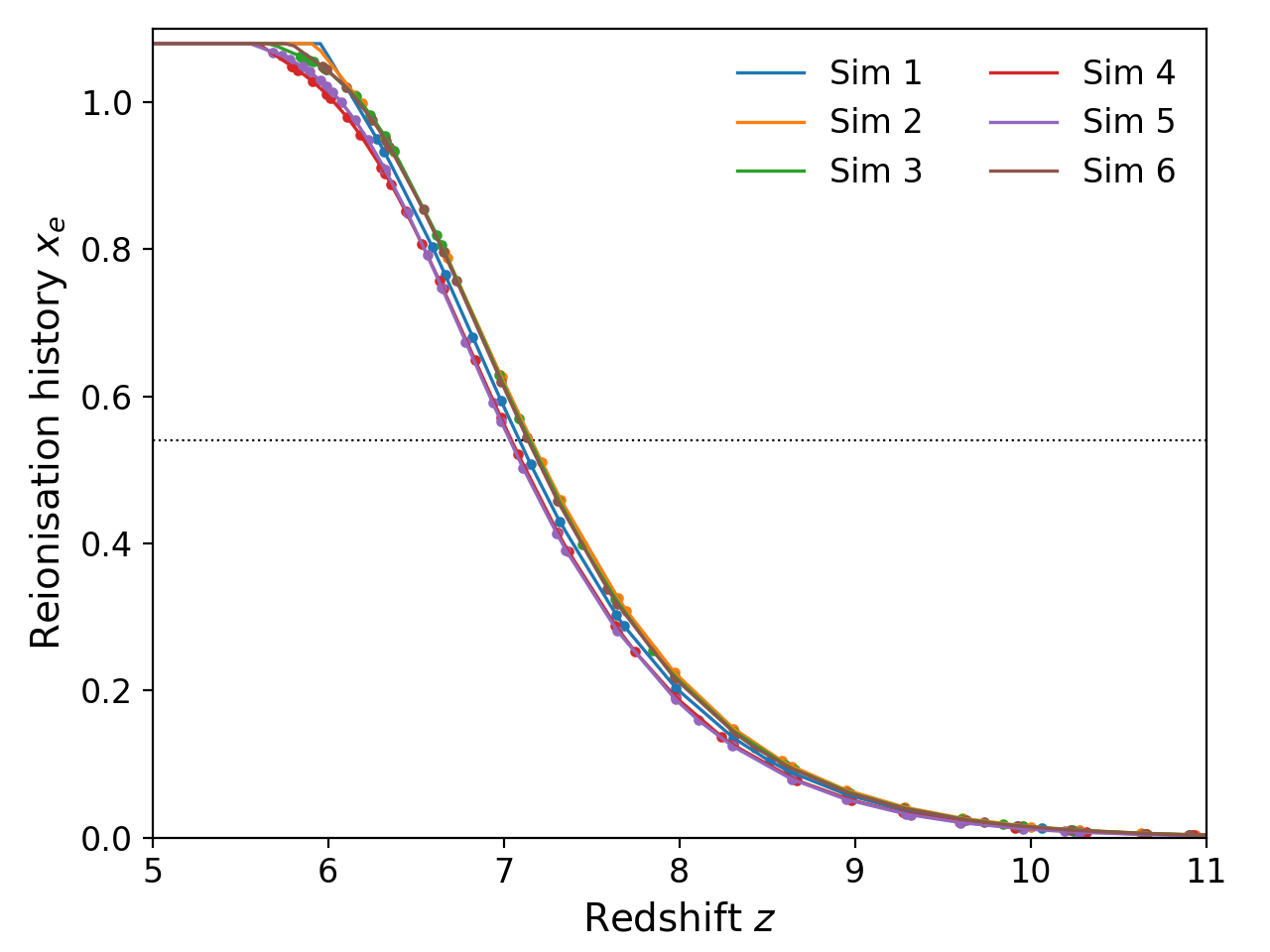}\\
    \includegraphics[width=0.8\columnwidth]{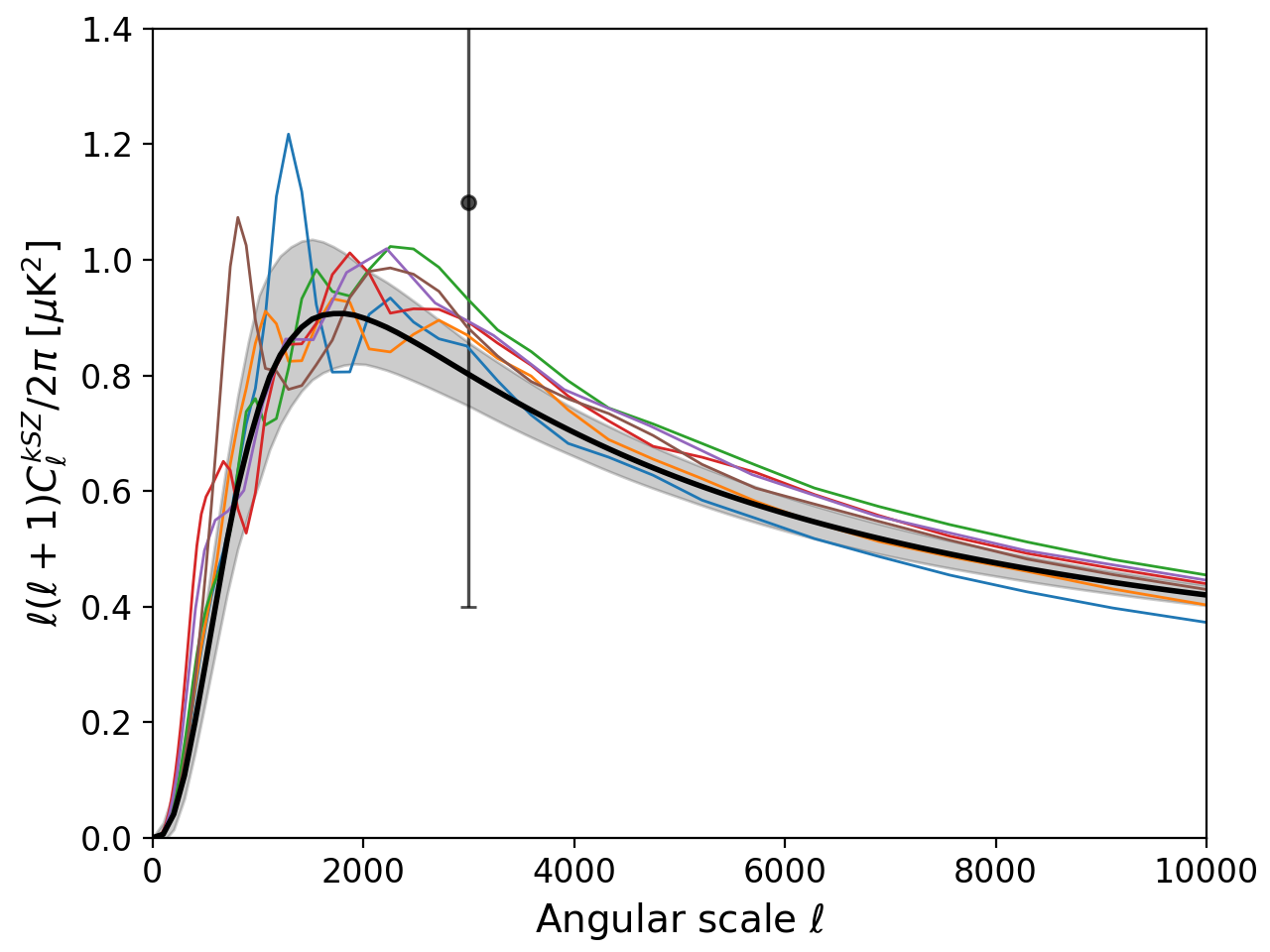}
    \caption{Results for our six simulations. \textit{Upper panel:} Global reionisation histories, for \ion{H}{II} and \ion{He}{II}. The dotted horizontal line marks the reionisation midpoint $\zrei$. \textit{Lower panel:} Angular kSZ power spectrum after fitting Eq.~\eqref{eq:full_Pee_expression} to the $\Pee(k,z)$ data points from our six simulations (thick solid line) compared to the spectra obtained when interpolating the data points for each simulation (thin solid lines). Error bars correspond to the propagation of the $68\%$ confidence interval on the fit parameters. The data point corresponds to constraints from \citet{SPT_2020} at $\ell=3000$.}
    \label{fig:ksz_six_sims}
\end{figure}

Now that we have a fitted $\Pee(k,z)$, we can compute the kSZ angular power spectrum using Eq.~\eqref{eq:def_C_ell_kSZ}. We find:
\begin{equation}
\dthree^\mathrm{p} = 0.80 \pm 0.06~\mu\mathrm{K}^2 
\end{equation}
and the angular scale at which the patchy angular spectrum reaches its maximum is $\lmax = 1800^{+300}_{-100}$. The angular patchy power spectrum is shown on the lower panel of Fig.~\ref{fig:ksz_six_sims}. The error bars correspond to the propagated $68\%$ confidence interval on the fit parameters. The amplitude of the homogeneous signal largely dominates that of the patchy signal, being about 4 times larger. The total kSZ amplitude reaches $\dthree = 4.2~\mu\mathrm{K}^2$ and so slightly exceeds the upper limits on the total kSZ amplitude given by SPT and Planck when SZxCIB correlations are allowed \citep[resp.][]{SPT_2020,planck_2016_reio} but is however within the error bars of the ACT results (\mbox{\citealt{sievers_2013_act}}).
With respect to the patchy signal, the amplitude is in perfect agreement with the claimed detection by the SPT at $\dthree^\mathrm{patchy} = 1.1 ^{+1.0}_{-0.7}~\mu\mathrm{K}^2$ \citep{SPT_2020}, noting that our simulations reionise in a time very close to their constraint $\Delta z= 1.1 ^{+1.6}_{-0.7}$. The spectrum exhibits the expected bump in amplitude, here around $\ell \sim 1800$, corresponding to larger scales than those found in other works \citep{iliev_2007,mesinger_2012_kSZ}, hinting at larger ionised bubbles on average. Fig.~\ref{fig:ksz_six_sims} gives an idea of the variance in the kSZ angular power spectrum for given physics -- in particular a given matter distribution, and very similar reionisation histories: the distribution among simulations gives a reionisation midpoint defined at $\zrei = 7.10 \pm 0.06$, corresponding to a range of kSZ power spectrum amplitude $\dthree = 0.80 \pm 0.06~\mu\mathrm{K}^2$ (at $68\%$ confidence level). Part of this variance can be related to sample variance, since our simulations have a too small side length ($L=128~\mathrm{Mpc}/h$) to avoid it \citep{iliev_2007}. We compare in Fig.~\ref{fig:ksz_six_sims} the kSZ power spectrum resulting from fitting Eq.~\eqref{eq:full_Pee_expression} on our six simulations simultaneously to the six spectra obtained when interpolating the $\Pee(k,z)$ data points available for each simulation: the six interpolated spectra lie withing the confidence limits of our best-fit.

\begin{figure}
    \centering
    \includegraphics[width=0.8\columnwidth]{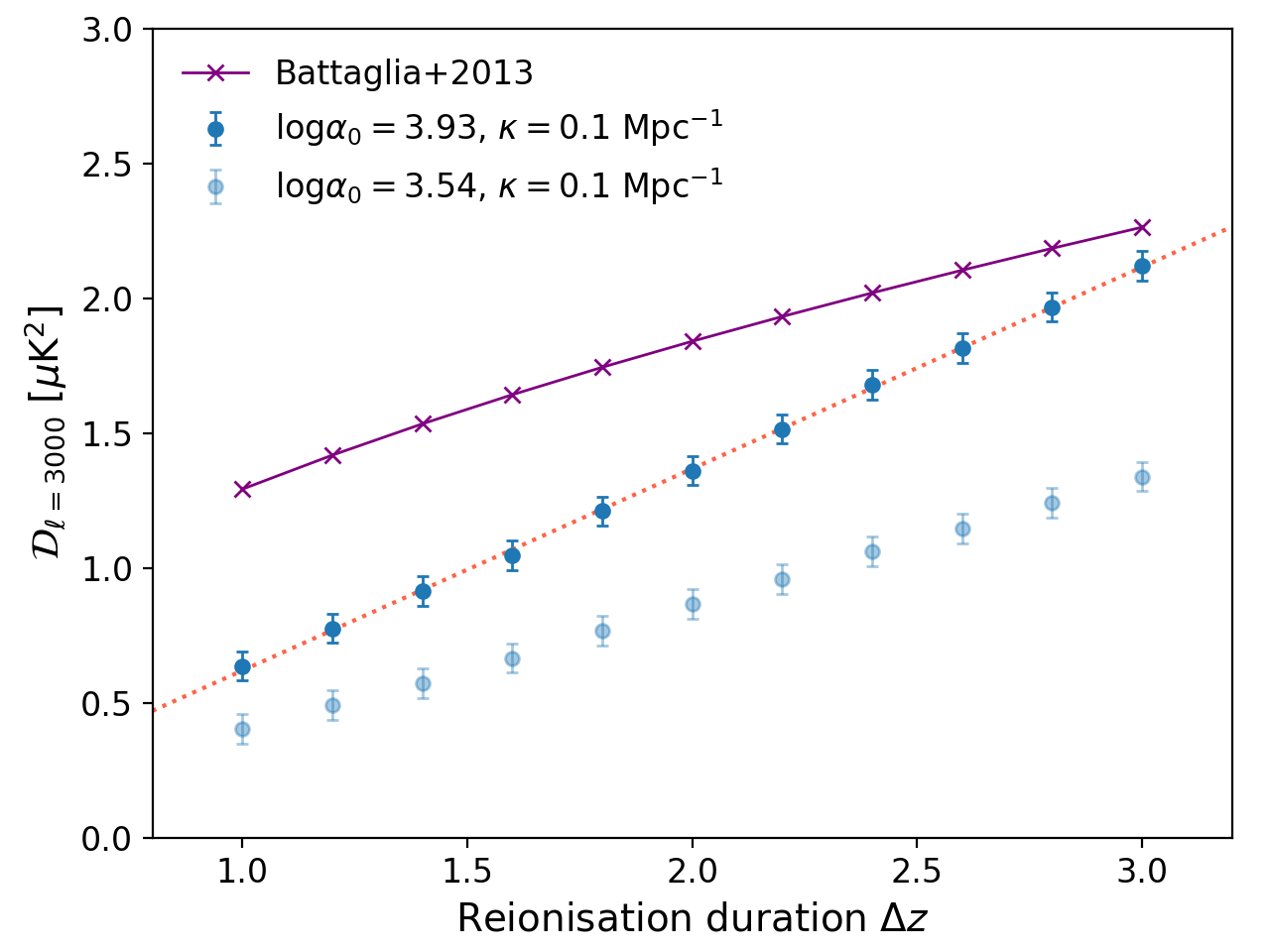}\\
    \includegraphics[width=0.8\columnwidth]{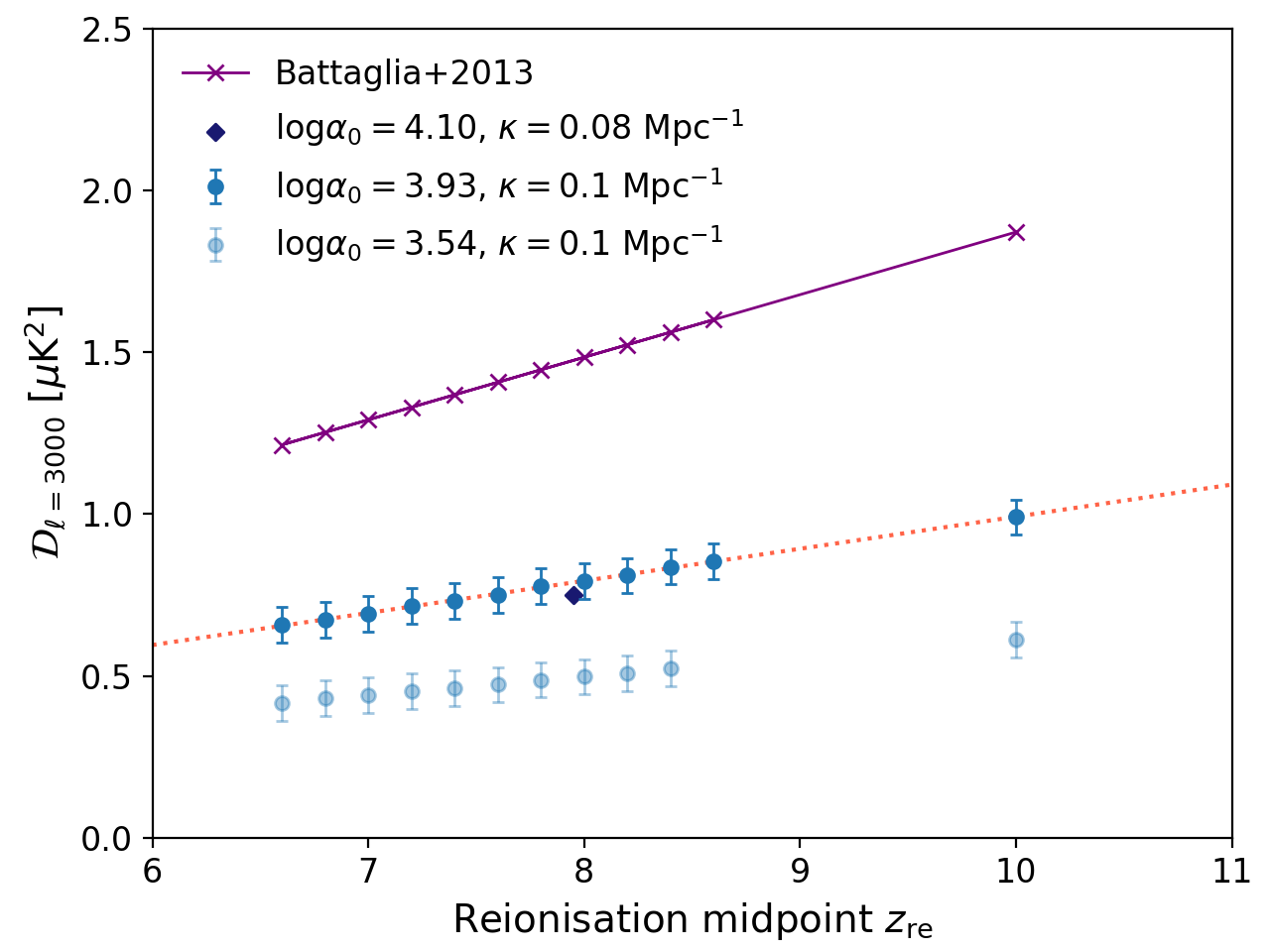}
    \caption{Evolution of the amplitude of the patchy power spectrum at $\ell = 3000$, $\dthree$ with the reionisation duration (upper panel) and the reionisation midpoint $\zrei$ (lower panel), for different values of our parameters. Error bars correspond to the dispersion of kSZ amplitude at $\ell=3000$ ($68\%$ confidence level) propagated from errors on the fit parameters. The diamond data point corresponds to a seventh simulation, with reionisation happening earlier. In both panels, results are compared to those of \cite{battaglia_2013_paperIII}, rescaled to our cosmology.}
    \label{fig:vary_xe_single_fit}
\end{figure}

\begin{figure*}
    \centering
    \includegraphics[width=.95\textwidth]{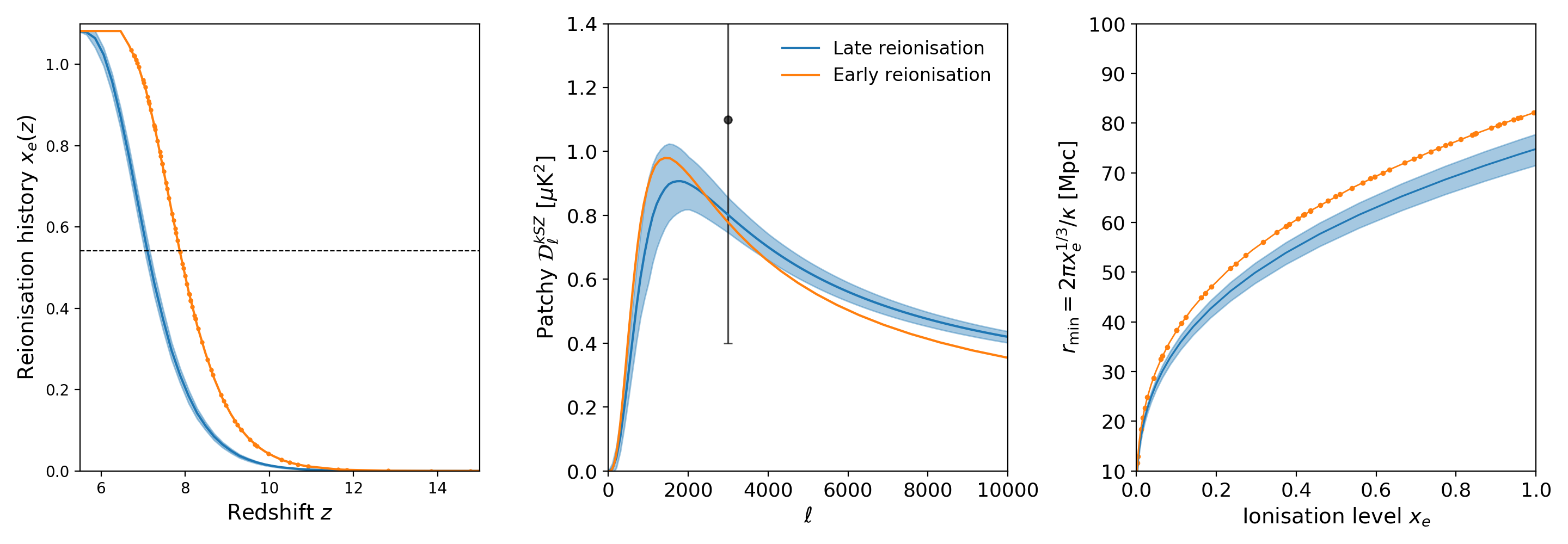}
    \caption{Comparison of results for our six initial simulations, corresponding to a late reionisation scenario, and for an additional seventh simulation, corresponding to an early reionisation scenario. \textit{Left panel:} Reionisation histories. \textit{Middle panel:} Patchy kSZ angular power spectra. The data point corresponds to constraints from \citet{SPT_2020}. \textit{Right panel:} Minimal size of ionised regions as a function of global ionised level. Shaded areas correspond to the $68\%$ confidence level on kSZ amplitude propagated from the probability distributions of the fit parameters.}
    \label{fig:earlier_reion_sim}
\end{figure*}

Fixing the fit parameters to their most likely value for the fourth simulation, we artificially vary the reionisation history and compute the corresponding power spectrum. We successively fix the reionisation redshift but increase its duration $\Delta z$ or fix the duration but shift the midpoint $\zrei$. This corresponds to a scenario where the reionisation morphology is exactly the same, but happens later or earlier in time. We find clear scaling relations between the amplitude of the signal at $\ell = 3000$, $\dthree$, and both the reionisation duration $\Delta z$ and its midpoint $\zrei$. However, they are sensibly different from the results of \citet{battaglia_2013_paperIII} as can be seen in Fig.~\ref{fig:vary_xe_single_fit}. Even after rescaling to their $\zrei = 8$ and cosmology, we get a much lower amplitude. Note also that their patchy spectra bump around $\ell = 3000$, whereas in our simulations the power has already dropped by $\ell=3000$ (Fig.~\ref{fig:ksz_six_sims}), hinting at a very different reionisation morphology from ours. When we vary $\kappa$ and $\alpha_0$ artificially, by fixing log$\alpha_0 = 3.54$ instead of 3.70 as before, there is still a scaling relation, but both the slope and the intercept change. All of this demonstrates that the amplitude of the patchy signal largely depends on the physics of reionisation (here via the $\kappa$ and $\alpha_0$ parameters) and $\Delta z$ and $\zrei$ are not sufficient to derive $\dthree$. Simulations closer to those used in \citet{battaglia_2013_paperIII} would likely give larger values for $\kappa$ and $\alpha_0$, therefore increasing the amplitude to values closer to the authors' results. To confirm this, we generate a new simulation, with same resolution and box size but with twice as much star formation as in the six initial simulations, therefore reionising earlier ($\zrei=7.94$) but on a similar redshift interval ($\Delta z = 1.20$). Applying the fitting procedure described above, we find log$\alpha_0 = 4.10$ and $\kappa=0.08~\mathrm{Mpc}^{-1}$. The resulting patchy kSZ power spectrum can be seen in Fig.~\ref{fig:earlier_reion_sim}, along with the reionisation histories and the evolution of the typical bubble size $r_\mathrm{min}=2\pi/\kappa x_e(z)^{1/3}$. Results for this simulation are compared with what was obtained for our six simulations. The kSZ spectrum corresponding to an early reionisation scenario bumps at larger scales ($\lmax=1400$) with a much larger maximum amplitude ($\mathcal{D}_\mathrm{max}=0.98~\mu\mathrm{K}^2$) but interestingly the amplitudes at $\ell=3000$ are similar. This suggests that focusing on $\dthree$ is not sufficient to characterise the kSZ signal.

 These results corroborate the work of \citet{park_2013}, who found that the scalings derived in \citet{battaglia_2013_paperIII} are largely dependent on the simulations they were calibrated on, and therefore cannot be used as a universal formula to constrain reionisation. Notably, an asymmetric reionisation history $x_e(z)$  naturally deviates from this relation. Global parameters such as $\Delta z$ and $\zrei$ are not sufficient to accurately describe the patchy kSZ signal, and one needs to take the physics of reionisation into account to get an accurate estimation of not only the shape, but also the amplitude of the power spectrum. Additionally, limiting ourselves to the amplitude at $\ell=3000$ to constrain reionisation can be misleading.

\subsection{Tests on other simulations}
\label{subsec:rsage}

\begin{figure*}
    \centering
    \includegraphics[width=.95\textwidth]{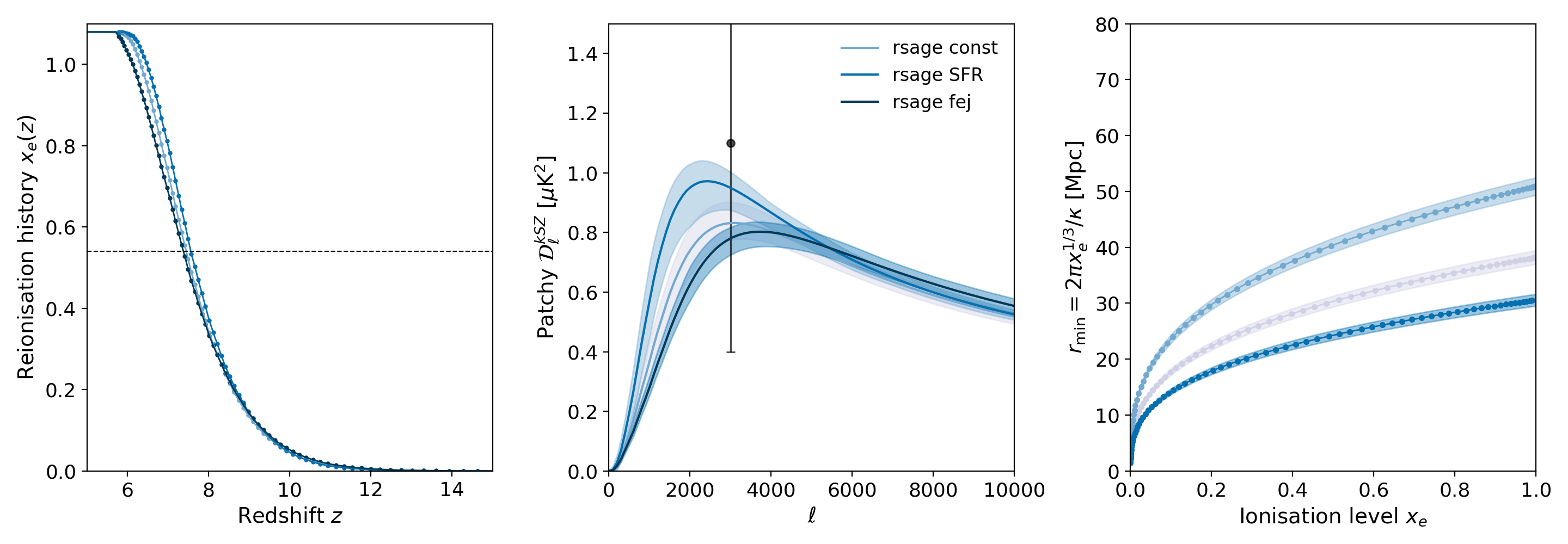}
    \includegraphics[width=.95\textwidth]{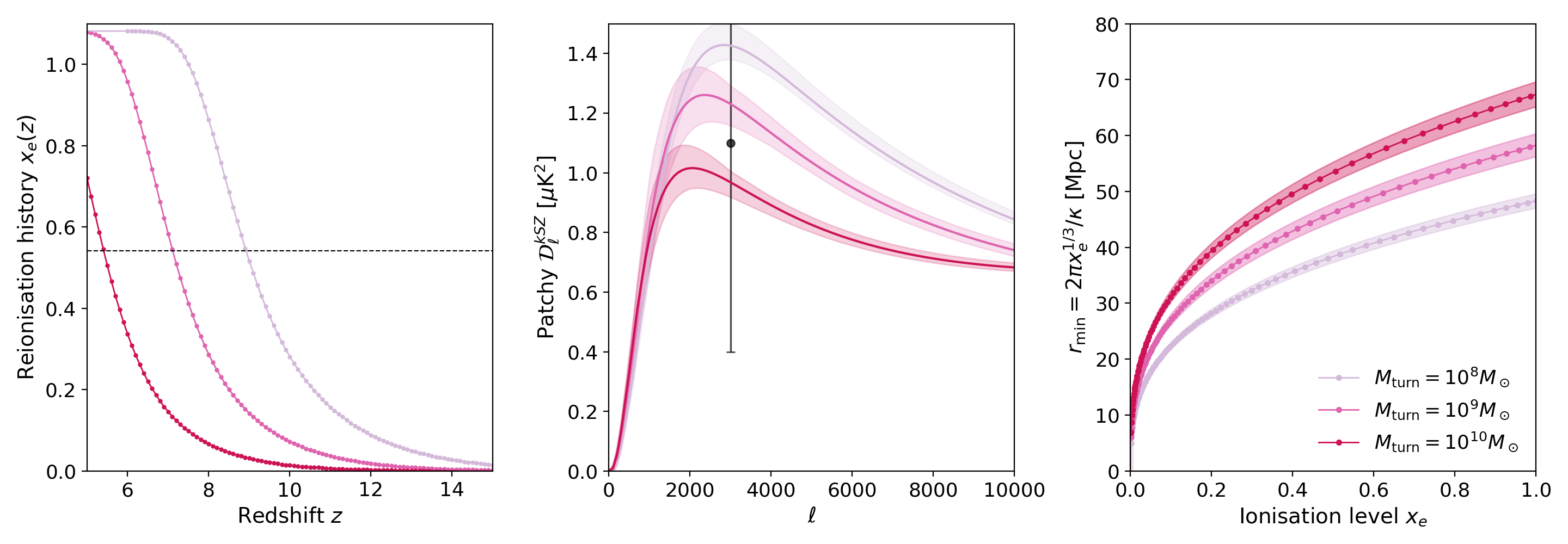}
    \caption{Comparison of results for the three \texttt{rsage} simulations (upper panels) and the three 21CMFAST runs (lower panels) considered. \textit{Left panels:} Reionisation histories. \textit{Middle panels:} Patchy kSZ angular power spectra. The data point corresponds to constraints from \citet{SPT_2020}. \textit{Right panels:} Minimal size of ionised regions as a function of global ionised level. The shaded areas correspond to the $68\%$ confidence interval propagated from the $68\%$ confidence intervals on the fit parameters.}
    \label{fig:results_ksz_rsage}
\end{figure*}

We now look at the \texttt{rsage} simulation, described in \citet{seiler_2019_fesc}, to test the robustness of our parameterisation. This simulation starts off as an $N$-body simulation \citep{seiler_2018}, containing $2400^3$ dark matter particles within a $160~\mathrm{Mpc}$ side box, resolving halos of mass $\sim 4 \times 10^8~\mathrm{M}_\odot$ with 32 particles.  Galaxies are evolved over cosmic time following the Semi-Analytic Galaxy Evolution (SAGE) model of \citet{croton_2016}, modified to include an improved modelling of galaxy evolution during the Epoch of Reionisation, including the feedback of ionisation on galaxy evolution. The semi-numerical code \texttt{cifog} \citep{cifog,scifog_anne} is used to generate an inhomogeneous ultraviolet background (UVB) and follow the evolution of ionised hydrogen during the EoR. Three versions of the \texttt{rsage} simulation are used, each corresponding to a different way of modelling the escape fraction $\fesc$ of ionising photons from their host galaxy into the IGM. The first, dubbed \texttt{rsage const}, takes $\fesc$ constant and equal to $20\%$. The second, \texttt{rsage fej}, considers a positive scaling of $\fesc$ with $f_\mathrm{ej}$, the fraction of baryons that have been ejected from the galaxy compared to the number remaining as hot and cold gas. In the last one, \texttt{rsage SFR}, $\fesc$ scales with the star formation rate and thus roughly with the halo mass. Because they are based on the same dark matter distribution, the three simulations start reionising at similar times ($z\sim13$), but different source properties lead to different reionisation histories, shown in the left upper panel of Fig.~\ref{fig:results_ksz_rsage}. In \texttt{rsage SFR}, the ionised bubbles are statistically larger than the other two simulations at a given redshift: this results into \texttt{rsage SFR} reaching $50\%$ of ionisation at $\zrei = 7.56$ vs. $\zrei = 7.45$ and $\zrei = 7.37$ for \texttt{rsage const} and \texttt{rsage fej} respectively, and the full ionisation being achieved in a shorter time. For more details, we refer the interested reader to \citet{seiler_2019_fesc}. Applying the fitting procedure to the three simulations, we find that the parameterisation of Eq.~\eqref{eq:full_Pee_expression} is an accurate description of the evolution of their $\Pee(k,z)$ spectra (detailed fit results are given in App.~\ref{subsec:app_rsage}). Resulting patchy kSZ angular power spectra are shown in the upper middle panel of Fig.~\ref{fig:results_ksz_rsage}. First, we find that \texttt{rsage fej} has the smallest $\alpha_0$ value, with $\log\alpha_0=2.87\pm0.04$. Because $\alpha_0$ is the maximum amplitude of the $\Pee(k,z)$ spectrum, built upon the free electrons density contrast field $\delta_e(r)= n_e(r)/\bar{n}_e - 1$, it will scale with the variance of the $n_e(r)$ field. Therefore a smaller $\alpha_0$ value is equivalent to a smaller field variance at all times. This is consistent with the picture of the different \texttt{rsage} simulations we have: as presented in \citet{seiler_2019_fesc}, \texttt{rsage fej} exhibits the smallest ionised bubbles on average. For a given filling fraction, a field made of many small bubbles covering the neutral background rather homogeneously will have smaller variance than one made of a few large bubbles. This in turn explains why \texttt{rsage SFR} gives the largest $\alpha_0$ value ($\log\alpha_0=3.47\pm0.04$), and, later, the largest kSZ amplitude (Fig.~\ref{fig:results_ksz_rsage}). Second, the \texttt{rsage SFR} simulation has the smallest value of $\kappa$ ($\kappa=0.123\pm0.004~\mathrm{Mpc}^{-1}$): the upper right panel of Fig.~\ref{fig:results_ksz_rsage} shows the evolution of $r_\mathrm{min}=2\pi/\kappa x_e^{1/3}$ with ionisation level for the three models. Because \texttt{rsage SFR} has the largest ionised bubbles on average \citep{seiler_2019_fesc}, this result confirms the interpretation of $1/\kappa$ as an estimate of the typical bubble size during reionisation. Additionally, the patchy power spectrum derived from \texttt{rsage SFR} peaks at larger angular scales ($\lmax\sim2400$) than for the other simulations, as can be seen in the upper middle panel of the figure. Interestingly, the largest $\alpha_0$ value leads to the strongest kSZ signal and the smallest $\kappa$ value to the spectrum whose bump is observed on the largest scales (the smallest $\lmax$). We investigate these potential links in the next section.

We now turn to three 21CMFAST \citep{21cmFAST_2007,21cmFAST_2011} simulations with dimensions $L=160~\mathrm{Mpc}$ for $256^3$ cells (same box size and resolution as \texttt{rsage}). Between the three runs, we vary the parameter $M_\mathrm{turn}$, the turnover mass, which corresponds to the minimum halo mass before exponential suppression of star formation.
For $M_\mathrm{turn} = 10^8 M_\odot$, the box is fully ionised by $\zend=6.25$ and the midpoint of reionisation is reached at $\zrei=8.92$ for a process lasting $\Delta z=1.91$. For $M_\mathrm{turn} = 10^9 M_\odot$, we find $\zend=4.69$, $\zrei=7.11$ and $\Delta z = 1.66$, which is closest to \texttt{rsage} and our initial six simulations. Finally, $M_\mathrm{turn} = 10^{10} M_\odot$ yields $\zend=3.37$,  $\zrei=5.41$ and $\Delta z = 1.47$. Indeed, the point of these simulations is not only to test the sensitivity of our approach to astrophysical parameters, but also to see the impact of very different reionisation histories on the patchy kSZ power. We find that Eq.~\eqref{eq:full_Pee_expression} again nicely fits the evolution of the $\Pee(k,z)$ spectra of these simulations, as shown in App.~\ref{subsec:app_21cmfast}. The resulting reionisation histories, patchy kSZ spectra and $r_\mathrm{min}(\xhii)$ are shown in the lower panels of Fig.~\ref{fig:results_ksz_rsage}. For $M_\mathrm{turn} = 10^8 M_\odot$, many small-mass halos are active ionising sources, resulting in an ionising field made of many small bubbles at the start of the process. This translates into this simulation having the largest best-fit $\kappa$ value of the three ($\kappa = 0.130\pm 0.003~\mathrm{Mpc}^{-1}$) and so the smallest $r_\mathrm{min}(\xhii)$. Naturally, the resulting kSZ spectrum peaks at smaller angular scales. For the other extreme case $M_\mathrm{turn} = 10^{10} M_\odot$, because the minimal mass required to start ionising is larger, the ionising sources are more efficient and the ionised bubbles larger. Indeed, we find a smaller value of $\kappa = 0.093\pm0.003~\mathrm{Mpc}^{-1}$. With larger bubbles, we also expect the variance in the ionisation field at the start of the process to be higher than if many small ionised regions cover the neutral background. This corresponds to the larger value of $\log \alpha_0 = 3.79 \pm 0.04$ found for this simulation, compared to $3.30 \pm 0.03$ for the first one. However, this larger value of $\alpha_0$ this time does not result into the strongest kSZ signal because of the very different reionisation histories of the three simulations. As we have seen in the previous section, the amplitude of the signal is strongly correlated with the duration and midpoint of reionisation, resulting in the first simulation ($M_\mathrm{turn} = 10^8 M_\odot$), corresponding to the earliest reionisation, having the strongest signal. This again emphasises how essential it is to consider both reionisation morphology and global history to derive the final kSZ spectrum.

These results show that our proposed simple two-parameter expression for $\Pee(k,z)$ can accurately describe different types of simulations, that is different types of physics, further validating the physical interpretation of the parameters $\alpha_0$ and $\kappa$ detailed in the next Section.

\section{Discussion and conclusions}
\label{sec:conclusion}

\subsection{Physical interpretation of the parameters}
\label{subsec:physical_interpretation}

\begin{figure}
    \centering
    \includegraphics[width=0.9\columnwidth]{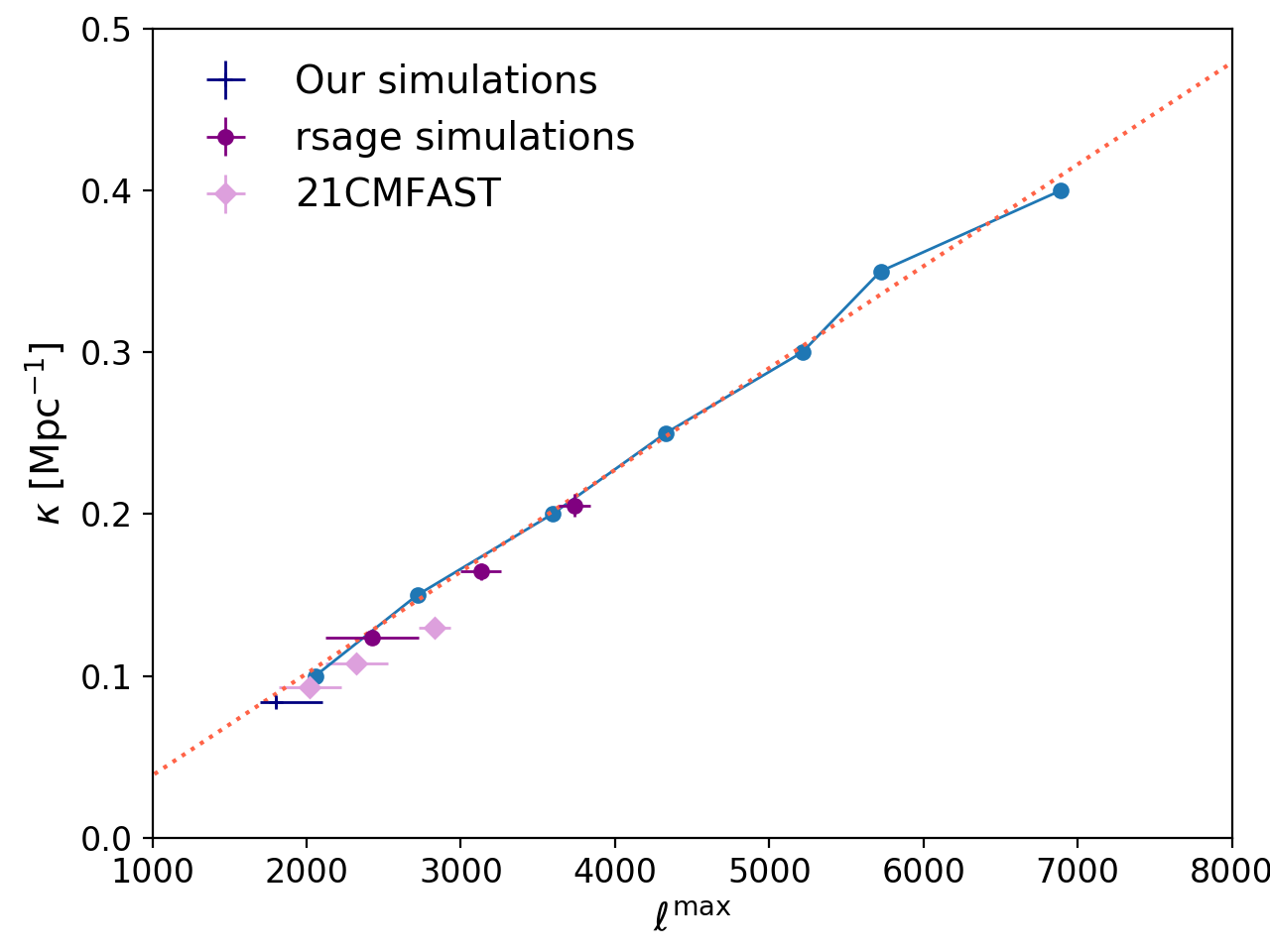}
    \caption{
    Evolution of the peaking angular scale of the patchy kSZ power spectrum for one given reionisation history but different values of the $\kappa$ parameter. The red dotted line is the result of a linear regression. Inferences are compared to results for different simulations.}
    \label{fig:kappa_rmin}
\end{figure}

Many previous works have empirically related the angular scale at which the patchy kSZ power spectrum reaches its maximum $\lmax$ to the typical size of bubbles during reionisation \citep{mcquinn_2005,iliev_2007,mesinger_2012_kSZ}. To test for this relation, we compute the patchy kSZ power spectrum for a given reionisation history $x_e(z)$ and $\alpha_0$ but let $\kappa$ values vary. We find a clear linear relation between $\kappa$ and $\lmax$ as shown in Fig.~\ref{fig:kappa_rmin}. Despite very different reionisation histories and physics at stake, previous results on the six high-resolution simulations, on 21CMFAST, and on \texttt{rsage}, roughly lie along this line. This means that a detection of the patchy power spectrum in CMB observations would make it possible to directly estimate $\lmax$, giving access to $\kappa$ without bias from reionisation history, and to the evolution of the typical bubble size. As the growth of ionised regions depends on the physical properties of early galaxies, such as their ionising efficiency or their star formation rate and on the density of the IGM, constraints on $\kappa$ could, in turn, give constraints on the nature of early light sources and of the early IGM.

Additionally, we can link the theoretical expression of the large-scale amplitude of the bubble power spectrum in Eq.~\eqref{eq:PS_toy_model} with our parameterisation of $\Pee(k,z)$ in Eq.~\eqref{eq:fit_formula}: $\alpha_0 x_e^{-1/5} \leftrightarrow 4/3\pi R^3/x_e(z)$. Because of the simplicity of the toy model, this relation is not an equivalence. For example, contrary to the toy model, in our simulations, the locations of the different ionised bubbles are correlated, following the underlying dark matter distribution and this correlation will add power to the spectrum on large scales. This analogy can however explain the correlation observed between $\alpha_0$ and $\kappa$ when fitting Eq.~\eqref{eq:full_Pee_expression} to data (recall that $R \propto 1/\kappa$). Finally, since $\alpha_0$ is independent of redshift, it will be a pre-factor for the left-hand side of Eq.~\eqref{eq:delta_Be}, therefore we expect a strong correlation between this parameter and the amplitude of the spectrum at $\ell=3000$ and with the maximum amplitude reached by the spectrum. We confirm this intuition by fixing the reionisation history and $\kappa$ but varying $\alpha_0$ on the range $3.0 < \log \alpha_0 < 4.4$ and comparing the resulting spectra: there is a clear linear relation between these two parameters and $\alpha_0$, but in this case results for \texttt{rsage} and 21CMFAST do not follow the correlation. Interestingly, the shape of the different resulting kSZ power spectra is strictly identical (namely, $\lmax$ does not change when varying $\alpha_0$), hinting at the fact that $\lmax$ depends only on $\kappa$ and not $\alpha_0$ or reionisation history. Therefore it will be possible to make an unbiased estimate of $\kappa$ from the shape of the measured spectrum. The \texttt{rsage} simulations show that, for a similar reionisation history, a larger value of $\alpha_0$ will lead to a stronger kSZ signal; but looking at 21CMFAST, we found that an early reionisation scenario can counterbalance this effect and lead to high amplitude despite low $\alpha_0$ values. This corroborates the results of \citet{mesinger_2012_kSZ}, which find that the amplitude of the spectrum is determined by both the morphology (and so the $\alpha_0$ value) and the reionisation history. Therefore, fitting CMB data to our parameterisation will likely lead to strongly correlated values of $\alpha_0$ and parameters such as $\Delta z$ or $\zrei$. Other methods should be used to constrain the reionisation history and break this degeneracy, such as constraints from the value of the Thomson optical depth, or astrophysical constraints on the IGM ionised level. Conversely, 21cm intensity mapping should be able to give independent constraints on $\alpha_0$.

\subsection{Conclusions \& prospects}

In this work, we have used state-of-the-art reionisation simulations \citep{aubert_2015_EMMA} to calibrate an analytical expression of the angular power spectrum of the kSZ effect stemming from patchy reionisation. We have shown that describing the shape, but also amplitude of the signal only in terms of global parameters such as the reionisation duration $\Delta z$ and its midpoint $\zrei$ is not sufficient: it is essential to take the physics of the process into account. In our new proposed expression, the parameters can be directly related to both the global reionisation history $x_e(z)$ and to the morphology of the process. With as few as these three parameters, we can fully recover the patchy kSZ angular power spectrum, in a way that is quick and easy to forward-model. Our formalism contrasts with current works, which use an arbitrary patchy kSZ power spectrum template enclosing an outdated model of reionisation. Applying it to CMB data will result in obtaining for the first time the actual shape of the patchy kSZ power spectrum, taking consistently into account reionisation history and morphology. In future works, we will apply this framework to CMB observations from SPT and, later, CMB-S4 experiments. Then, the inferred values of $\alpha_0$ and $\kappa$ will provide us with detailed information about the physics of reionisation: $\kappa$ will constrain the growth of ionised bubbles with time and $\alpha_0$ the evolution of the variance of the ionisation field during EoR, both being related to the ionising properties of early galaxies. The complex derivation of the kSZ signal, based on a series of integrals, leads to correlations between our parameters. For example, a high amplitude of the spectrum can be explained either by a large value of $\alpha_0$ due to a high ionising efficiency of galaxies, or by an early reionisation. Such degeneracies, however, could be broken by combining CMB data with other observations: astrophysical observations of early galaxies and quasars will help grasp the global history of reionisation and constrain parameters such as $\Delta z$ and $\zrei$, while 21cm intensity mapping will help understand reionisation morphology, putting independent constraints on $\alpha_0$ and $\kappa$. The main challenge remains to separate first the kSZ signal from other foregrounds, and then the patchy kSZ signal from the homogeneous one. To solve the first part of this problem, \citet{Calabrese_2014} suggest to subtract the theoretical primary power spectrum (derived from independent cosmological parameter constraints obtained from polarisation measurements) from the observed one so that the signal left is the kSZ power spectrum alone. Secondly, one would need a good description of the homogeneous spectrum, similar to the results of \citet{shaw_2012} but updated with more recent simulations, in order to estimate how accurately one can recover the patchy signal.
Additionally, this result sheds light on the scaling relations observed in previous works by giving them a physical ground. For example, features in the free electron contrast density power spectrum explain the relation between the amplitude at which the patchy kSZ spectrum bumps $\lmax$ and the typical bubble size, which was observed empirically in many previous works \citep{mcquinn_2005,iliev_2007,mesinger_2012_kSZ}.

On average, our results are in good agreement with previous works, despite a low amplitude of the patchy kSZ angular power spectrum at $\ell=3000$ ($\sim 0.80~\mu\mathrm{K}^2$) for our fiducial simulations. There is undoubtedly a bump around scales $\ell \sim 2000$ that can be related to the typical bubble size and the amplitude of the total (patchy) kinetic SZ spectrum ranges from 4 to $5~\mu\mathrm{K}^2$ (0.5 to $1.5~\mu\mathrm{K}^2$, respectively) for plausible reionisation scenarios, therefore lying within the error bars of the latest observational results of ACT \mbox{\citep{sievers_2013_act}} and SPT \citep{SPT_2020}. We have found that the majority of the patchy kSZ signal stems from scales $10^{-3} < k / \mathrm{Mpc}^{-1} < 1$ and from the core of the reionisation process ($10\% < x_e < 80\%$), ranges on which we must focus our efforts to obtain an accurate description. This analysis does not consider third- and fourth-order components of the kSZ signal, which can represent as much as $10\%$ of the total signal \citep{alvarez_2016}, and uses a coarse approximation for the electrons density - velocity cross spectra. In contrast to previous works, these results are not simulation-dependent as we have tested the robustness of our model by confronting it to different types of simulations, capturing different aspects of the process. However, the analytic formulation of our derivations was calibrated on a relatively small simulation, of side length $128~\mathrm{Mpc}/h$, which could bias our results. To further support our approach, using a larger radiative hydrodynamical simulation would be useful. Additionally, one could derive the kSZ power from lightcones constructed with our simulation, but the limited size of the simulation might lead to a significant underestimate of the kSZ power \citep{shaw_2012,alvarez_2016}.

\begin{acknowledgements}

The authors thank Anne Hutter and Jacob Seiler for kindly providing runs of their simulations, as well as Jonathan Pritchard and Ian Hothi for fruitful discussions at various stages of this analysis. They also thank the referee for useful comments which helped imoprove the quality of these results. AG acknowledges financial support from the European Research Council under ERC grant number 638743-FIRSTDAWN and her work is supported by a PhD studentship from the UK Science and Technology Facilities Council (STFC). This work was initiated during LSS2LSS, a $\Psi^2$ thematic programme organised by the Institut d'Astrophysique Spatiale and funded by the Université Paris-Saclay in July 2018 (see \url{https://www.ias.u-psud.fr/LSS2LSS}). The authors were granted access to the HPC resources of CINES and IDRIS under the allocation A0070411049 attributed by GENCI (Grand Equipement National de Calcul Intensif) and the Jean-Zay Grand Challenge (CT4) "Émulation de simulations de Réionisation par apprentissage profond". This work was additionally supported by the Programme National Cosmology et Galaxies (PNCG) of CNRS/INSU with INP and IN2P3, co-funded by CEA and CNES. SI was supported by the European Structural and Investment Fund and the Czech Ministry of Education, Youth and Sports (Project CoGraDS - CZ.02.1.01/0.0/0.0/15\_003/0000437).\\

This research made use of \texttt{astropy}, a community-developed core Python package for astronomy \citep{astropy,astropy2}; \texttt{matplotlib}, a Python library for publication quality graphics \citep{hunter_2007}; \texttt{scipy}, a Python-based ecosystem of open-source software for mathematics, science, and engineering \citep{scipy} -- including \texttt{numpy} \citep{numpy}, and \texttt{emcee}, an implementation of the affine invariant MCMC ensemble sampler \citep{emcee}.
 
\end{acknowledgements}

\bibliographystyle{aa} 
\bibliography{biblio} 

\appendix




\section{Variations on the fit}
\label{app:cov_matrices}

\subsection{Six fits for six simulations}

\begin{figure}
    \centering
    \includegraphics[width=\columnwidth]{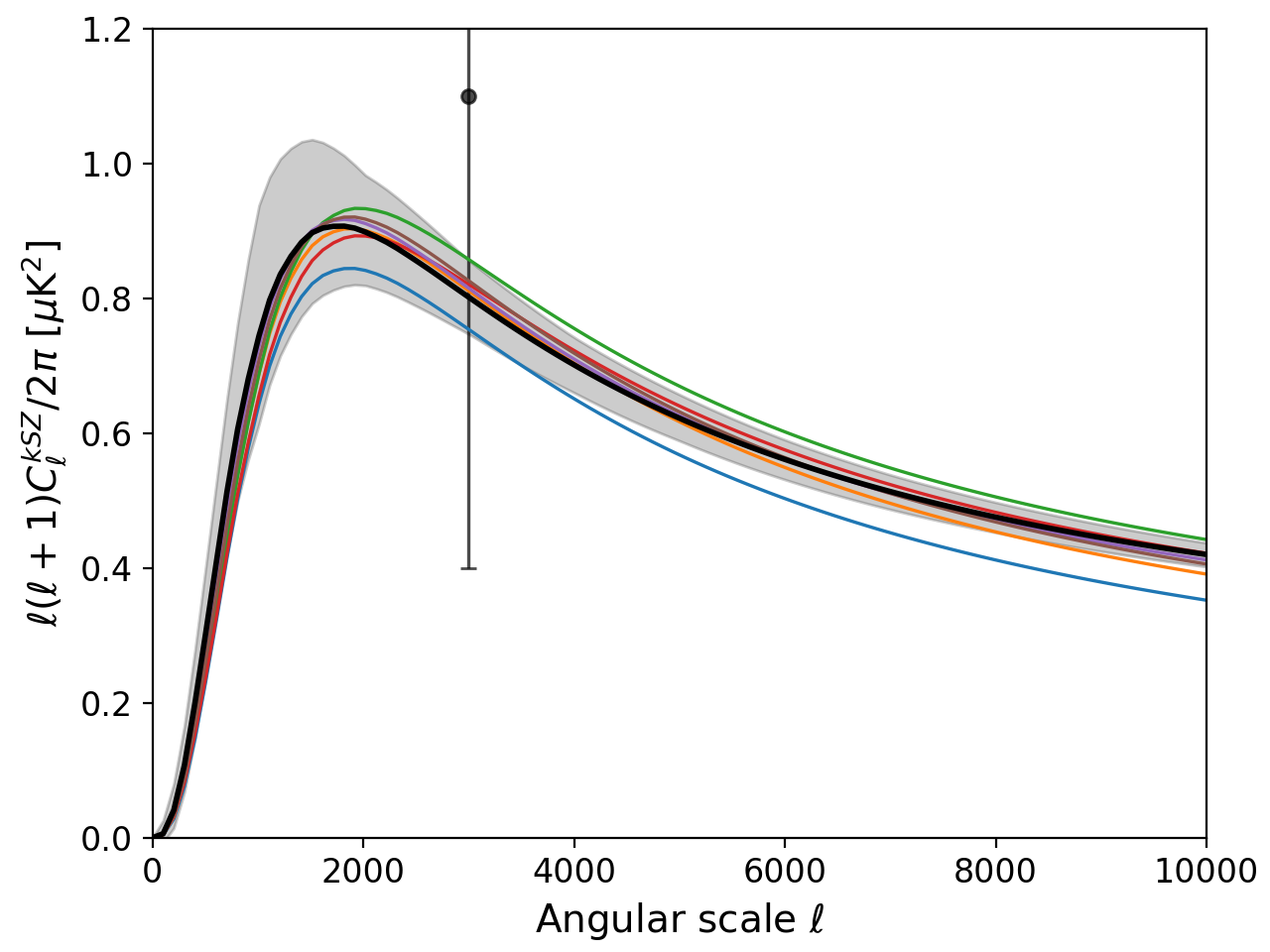}
    \caption{Comparison of the patchy kSZ power spectra resulting from one fit on the six simulations (black solid line, with $68\%$ confidence interval as the shaded area) or from six fits (coloured solid lines). The data point corresponds to constraints from \citet{SPT_2020}.}
    \label{fig:app_six_fits}
\end{figure}

Instead of fitting the six simulations simultaneously, we choose to fit each simulation individually to Eq.~\ref{eq:full_Pee_expression} with the same error bars as the fitting procedure described in Sec.~\ref{sec:results}. This allows to use the original $\Pee(k,z)$ data points from each simulation, without interpolating them, and the original reionisation history rather than an averaged one. The results are shown in Table \ref{tab:six_fits}, where the maximum likelihood parameters, along with their $68\%$ confidence intervals, and the corresponding values of $\dthree$ and $\lmax$ are given. The six maximum likelihood values of $\alpha_0$ and $\kappa$ lie within the $95\%$ confidence interval of the parameter distributions obtained in  Sec.~\ref{sec:results} and so do the resulting patchy kSZ spectra, as shown in Fig.~\ref{fig:app_six_fits}.

\begin{table}
    \caption{Results obtained when fitting Eq.~\ref{eq:full_Pee_expression} to the six simulations separately. Maximum likelihood parameters are given with 68\% confidence intervals. }
    \label{tab:six_fits}
    \centering
    \begin{tabular}{c|ccccccccc}
           Sim &  $\log\alpha_0$ [Mpc$^3$] & $\kappa$ [Mpc$^{-1}$] & $\mathcal{D}^p_{3000}$ [$\mu$K$^2$]& $\lmax$\\
        \hline
           \hline
         1 & $3.86 \pm 0.08$ & $0.093 \pm 0.006$ & 0.75 $\mu$K$^2$  & 1900\\
          2 & $3.85 \pm 0.08$ & $0.094 \pm 0.006$ & 0.81 $\mu$K$^2$ & 1900\\
          3 & $3.80 \pm 0.08$ & $0.098 \pm 0.007$ & 0.86 $\mu$K$^2$ & 1900\\
         4 & $3.78 \pm 0.08$ & $0.100 \pm 0.007$ & 0.82 $\mu$K$^2$ & 1900\\
         5 & $3.91 \pm 0.08$ & $0.089 \pm 0.007$ & 0.82 $\mu$K$^2$ & 1800\\
         6 & $3.87 \pm 0.08$ & $0.093 \pm 0.006$ & 0.83 $\mu$K$^2$  & 1900\\
    \end{tabular}
\end{table}

\subsection{Attempt at deriving a covariance matrix from a sample of six}

Because of the very insufficient number of simulations available to derive a covariance matrix, even when bootstrapping, we choose to average covariance matrices over bins.

\paragraph{Average over $z$-bins}
First, we choose to ignore correlations between scales over redshifts and use a covariance matrix $\bm{C}$, average of the $6 \times 10$ covariance matrices obtained for each simulation and each redshift bin. $\bm{C}$ has therefore dimensions $(20,20)$\footnote{Recall we have 10 redshift bins and 20 scale bins after interpolating the spectra.}. We fit Eq.~\ref{eq:full_Pee_expression} to the six simulations, trying to minimise:
\begin{equation}
    \chi^2 = \sum_{z_i} X_i^\mathrm{T}\, \bm{C}^{-1}\, X_i,
\end{equation}
where $X_i=\Pee^\mathrm{data}(\{k_j\},x_i) - \Pee^\mathrm{model}(\{k_j\},x_i) $. We find a minimal reduced $\chi^2$ of 125, reached for $\log \alpha_0 = 4.12$ and $\kappa = 0.078~\mathrm{Mpc}^{-1}$ and giving $\dthree^\mathrm{p} = 0.97~\mu\mathrm{K}^2$ and $\lmax=1500$. This difference comes from a poor match between the maximum likelihood $\Pee(k,z)$ and the data points on scales $0.1 < k/\mathrm{Mpc}^{-1} < 0.3$. These scales correspond to the power cut-off, so that the value of $\kappa$ is poorly constrained and, later, the kSZ power spectrum is distorted.

\paragraph{Average over $k$-bins}
Secondly, we choose to ignore correlations between redshifts over scales and use a covariance matrix $\bm{C}$, average of the $6 \times 20$ covariance matrices obtained for each simulation and each scale bin. $\bm{C}$ has therefore dimensions $(10,10)$. Comparing the correlation coefficients obtained for the two approaches, we note that the correlations are higher for this approach. We fit Eq.~\ref{eq:full_Pee_expression} to the six simulations, trying to minimise:
\begin{equation}
    \chi^2 = \sum_{k_i} X_i^\mathrm{T}\, \bm{C}^{-1}\, X_i,
\end{equation}
where $X_i=\Pee^\mathrm{data}(k_i,\{x_j\}) - \Pee^\mathrm{model}(k_i,\{x_j\}) $.
We find a minimal reduced $\chi^2$ of 4.46, reached for $\log \alpha_0 = 3.65$ and $\kappa = 0.135~\mathrm{Mpc}^{-1}$ and giving $\dthree^\mathrm{p} = 1.46~\mu\mathrm{K}^2$ and $\lmax=2700$. The excess power comes from the fact that the fit systematically overestimate the $\Pee$ power on small scales ($k>0.3~\mathrm{Mpc}^{-1}$).

\section{Detailed results on \texttt{rsage} and 21CMFAST}
\label{app:details_other_sims}

\begin{figure*}
    \centering  \includegraphics[width=.9\textwidth]{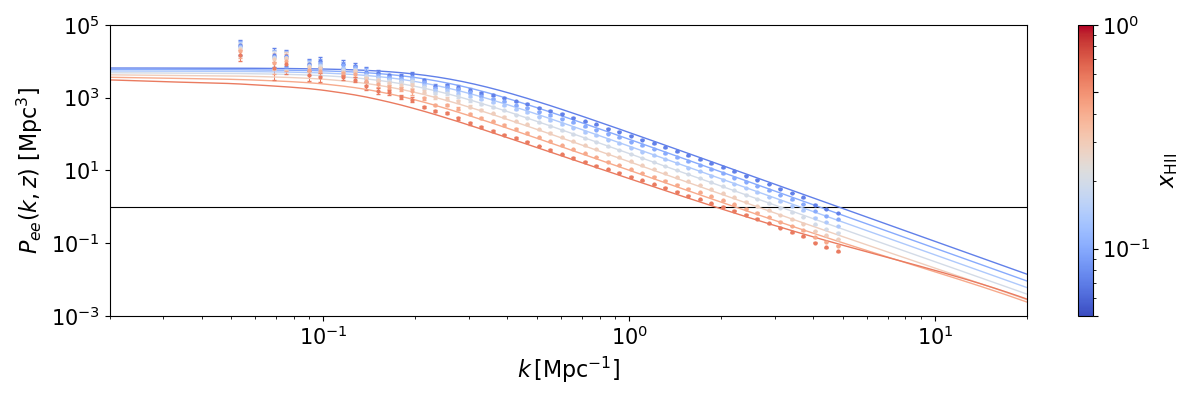}
    \includegraphics[width=.9\textwidth]{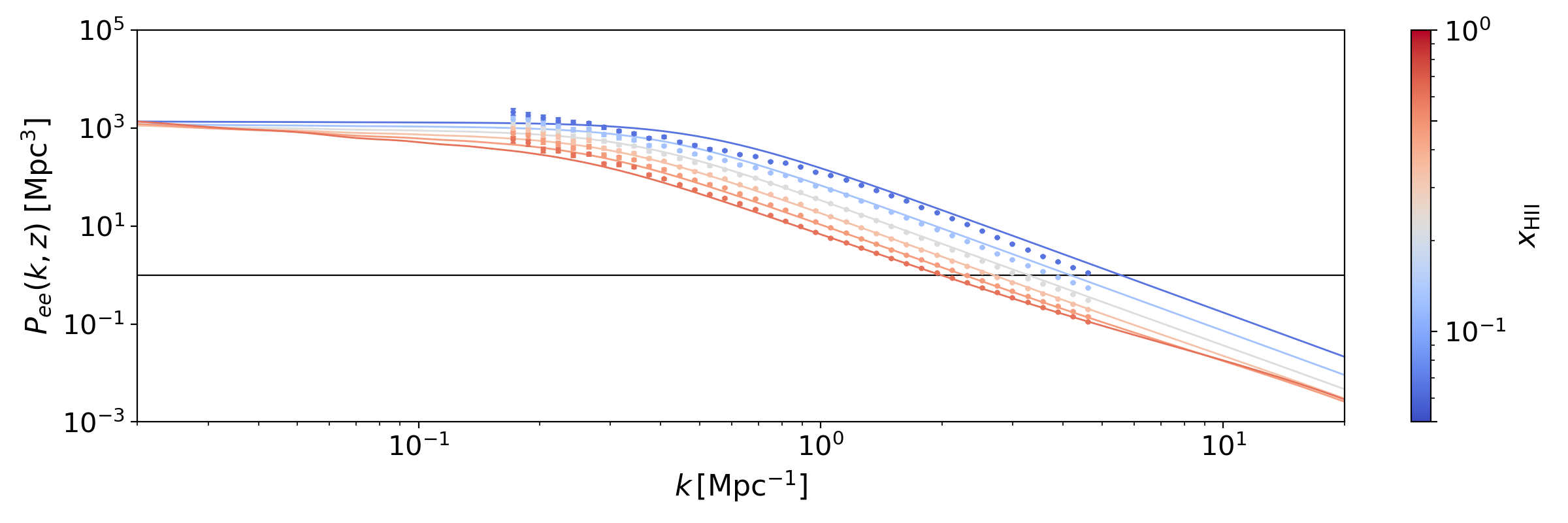}  
    \caption{Result of fitting Eq,~\eqref{eq:fit_formula} to the spectra of the 21CMFAST run for $M_\mathrm{turn}=10^9M_\odot$ (upper panel) and of \texttt{rsage fej} (lower panel). The error bars correspond to the $68\%$ confidence level on the spectra of 20 realisations of the same 21CMFAST run.}
    \label{fig:app_Pee_vs_k}
\end{figure*}

\subsection{Fits on 21CMFAST}
\label{subsec:app_21cmfast}

We now fit Eq.~\eqref{eq:full_Pee_expression} to the power spectra of our three 21CMFAST runs. To account for sample variance, we perform 20 realisations of each simulation -- the choice of 20 being motivated by \citet{kaur_202} and computational limitations. From these 20 realisations we derive relative error bars on $\Pee(k,z)$ values, corresponding to the $68\%$ confidence level on the distribution of values for each bin. The results obtained for 21CMFAST and their interpretation are consistent with what is obtained for the other simulations. The upper panel of Fig.~\ref{fig:app_Pee_vs_k} shows the best-fit model for $\Pee(k,z)$, along with snapshot values, for the second simulation.
    
\subsection{Fits on \texttt{rsage}}
\label{subsec:app_rsage}

Because we only have one realisation of each \texttt{rsage} simulation, we apply the relative error bars derived from 21CMFAST to the \texttt{rsage} $\Pee(k,z)$ data points. On the scales and redshifts range covered by the fit, the error bars $\sigma(k,z)$ derived from the 20 realisations of each of the three 21CMFAST simulations follow $\sigma(k,z) = 10^b Pee(k,z) k^a $, where $a = -1.12 \pm 0.79$ and $b=-1.74\pm0.70$ have been found by fitting the $\sigma(k,z)$ values of the 60 simulations simultaneously. We then apply this expression to the spectra of the \texttt{rsage} simulations, a reasonable first approximation of cosmic variance. We fit Eq.~\eqref{eq:full_Pee_expression} to the spectra of the three simulations. The lower panel of Fig.~\ref{fig:app_Pee_vs_k} shows the best-fit model for $\Pee(k,z)$, along with snapshot values, for \texttt{rsage fej}. Note that here, we only show the spectra on the redshift range used for the fit, where the power-law structure is not as striking as for higher redshifts.

\end{document}